\documentclass[twocolumn,times,trackchanges, dvipsnames]{aastex631}
\usepackage[utf8]{inputenc}

\begin{document}
\title{A Dust-Trapping Ring in the Planet-Hosting Disk of Elias\,2-24}

\author{Adolfo S. Carvalho}
\affiliation{Department of Astronomy; California Institute of Technology; Pasadena, CA 91125, USA}
\author{Laura M.\ Pérez}
\affiliation{Departamento de Astronomía; Universidad de Chile; Camino El Observatorio 1515, Las Condes, Santiago, Chile}
\author{Anibal Sierra}
\affiliation{Departamento de Astronomía; Universidad de Chile; Camino El Observatorio 1515, Las Condes, Santiago, Chile}
\author{Maria Jesus Mellado}
\affiliation{Departamento de Astronomía; Universidad de Chile; Camino El Observatorio 1515, Las Condes, Santiago, Chile}
\author{Lynne A. Hillenbrand}
\affiliation{Department of Astronomy; California Institute of Technology; Pasadena, CA 91125, USA}
\author{Sean Andrews}
\affiliation{Center for Astrophysics; Harvard \& Smithsonian; Cambridge, MA 02138, USA}
\author{Myriam Benisty}
\affiliation{Univ. Grenoble Alpes, CNRS, IPAG, 38000 Grenoble, France}
\affiliation{Université Côte d'Azur, Observatoire de la Côte d'Azur, CNRS, Laboratoire Lagrange, France}
\author{Tilman Birnstiel}
\affiliation{University Observatory, Faculty of Physics, Ludwig-Maximilians-Universität München, Scheinerstr. 1, 81679 Munich, Germany}
\affiliation{Exzellenzcluster ORIGINS, Boltzmannstr. 2, D-85748 Garching, Germany}
\author{John M. Carpenter}
\affiliation{Joint ALMA Observatory, Avenida Alonso de Córdova 3107, Vitacura, Santiago, Chile}
\author{Viviana V. Guzmán}
\affiliation{Instituto de Astrofísica, Pontificia Universidad Católica de Chile, Av. Vicuña Mackenna 4860, 7820436 Macul, Santiago, Chile}
\author{Jane Huang}
\affiliation{Department of Astronomy, Columbia University, 538 W. 120th Street, Pupin Hall, New York, NY, USA}
\author{Andrea Isella}
\affiliation{Department of Physics and Astronomy, Rice University 6100 Main Street, MS-108, Houston, TX 77005, USA}
\author{Nicolas Kurtovic}
\affiliation{Max-Planck-Institut für Astronomie, Königstuhl 17, 69117, Heidelberg, Germany}
\author{Luca Ricci}
\affiliation{Department of Physics and Astronomy, California State University Northridge, 18111 Nordhoff Street, Northridge, CA 91330, USA}
\author{David J. Wilner}
\affiliation{Center for Astrophysics; Harvard \& Smithsonian; Cambridge, MA 02138, USA}

\begin{abstract}
    Rings and gaps are among the most widely observed forms of substructure in protoplanetary disks. A gap-ring pair may be formed when a planet carves a gap in the disk, which produces a local pressure maximum following the gap that traps inwardly drifting dust grains and appears as a bright ring due to the enhanced dust density. A dust-trapping ring would provide a promising environment for solid growth and possibly planetesimal production via the streaming instability. We present evidence of dust trapping in the bright ring of the planet-hosting disk Elias\,2-24, from the analysis of 1.3 mm and 3 mm ALMA observations at high spatial resolution (0.029 arcsec, 4.0 au). We leverage the high spatial resolution to demonstrate that larger grains are more efficiently trapped and place constraints on the local turbulence ($8 \times 10^{-4} < \alpha_\mathrm{turb} < 0.03$) and the gas-to-dust ratio ($\Sigma_g / \Sigma_d < 30$) in the ring. Using a scattering-included marginal probability analysis we measure a total dust disk mass of $M_\mathrm{dust} = 13.8^{+0.7}_{-0.5} \times 10^{-4} \ M_\odot$. We also show that at the orbital radius of the proposed perturber, the gap is cleared of material down to a flux contrast of 10$^{-3}$ of the peak flux in the disk.
\end{abstract}

\section{Introduction}
The earliest observations at high spatial resolution from the Atacama Large Millimeter/submillimeter Array (ALMA) showed that there is significant substructure in the dust distribution of disks \citep[e.g.,][]{ALMA_HLTau_2015ApJ, Andrews_DSHARP_I_2018ApJ}, such as dark gaps and bright rings, wherein the dust surface density may be much lower (gaps) or higher (rings) than the typically adopted smooth (power-law) profile. 

The origin of substructures is a topic of ongoing debate. One possibility is that dark gaps are caused by young massive planets, clearing out gas and dust as they orbit their central star \citep[see][for a review of this and other proposed mechanisms]{Bae_substructure_ppvii_2022arXiv221013314B}. In this scenario, the cleared-out gap would produce a region where the local pressure decreases rapidly in the gap, then increases toward the outer edge of the gap as the gas surface density rises again. This pressure gradient inversion would act against the direction of dust migration \citep[for small grains, this is driven by the overall negative gas pressure gradient in the disk,][]{Birnstiel_review_araa_2023arXiv231213287B} and trap migrating dust grains on the pressure maxima \citep[see, e.g.,][]{pinilla_trapping_2012A&A}. Over time, the surface density of solids on the outer edge of the gap would increase as they pile up, producing a bright ring of millimeter continuum emission \citep{Testi_dustEvolutionInPPDs_2014prpl}. 

The ability of gap/ring pairs to form dust traps in disks is an avenue by which the radial migration of larger dust grains can be slowed sufficiently, to allow the grains to grow rapidly into pebbles via collisions \citep[e.g.,][]{drazkowska-graingrowth-2019ApJ}. Due to the increased concentration of pebbles around the trap, the streaming instability can be triggered \citep{YoudinGoodman_streamingInstability_2005ApJ, Johansen_StreamingInstabilityPlanetesimalFormation_2007Natur, Lesur_HydroDynamicsOfPPDs_PPVII_2023ASPC} and the clumps of pebbles may gravitationally collapse into planetesimals. 

Validating that gap/ring pairs represent bonafide dust traps requires a clear measurement that larger particles are relatively more concentrated in the ring than smaller particles. This can be accomplished with multiwavelength observations, because the larger particles emit preferentially at longer wavelengths. Although the sample of systems targeted for this type of study is growing \citep[see e.g.,][]{ALMA_HLTau_2015ApJ,Huang_TwHya_Multiband_2018ApJ,carrasco-gonzalez_HLTauDustDist_2019ApJ,Huang_GMAur_Multiband_2020ApJ,Macias_TWHya_2021A&A, Guidi_HD163296_2022A&A}, the need for extremely high sensitivity and spatial resolution data limits us to the brightest and/or most massive disks.




A promising system -- with a gap/ring pair in the disk and an embedded protoplanet candidate -- is Elias\,2-24 (henceforth Elias\,24) located at J2000 RA = 16h\,26m\,24.08s, Dec = -24d\,16m\,13.9s, and a distance of $139 \pm 1$ pc \citep{gaia_DR3_2023A&A}. 
ALMA observations at 1.3\,mm revealed a deep gap centered at a radius of 58 au and a bright ring at 78 au \citep{Cieza_Elias24_gap_2017ApJ, dipierro_Elias24_gap_2018MNRAS,Huang_DSHARPII_2018ApJ}. 
A planetary-mass companion was discovered inside the gap at a separation of 57 au (0.411$^{\prime\prime}$), using VLT/NaCo $L^\prime$ diffraction-limited imaging \citep{Jorquera_Elias24_CPD_2021AJ}. A counterpart traced in $^{12}$CO was found coincident to the $L^\prime$ point-source and detached in velocity space from the Keplerian disk motion \citep{pinte_Elias24_CO_2023MNRAS}, making Elias\,24 is the first system where a candidate detection in kinematics and direct emission agree, strengthening the companion hypothesis.

\begin{figure*}[!t]
    \centering
    \includegraphics[width=0.49\linewidth]{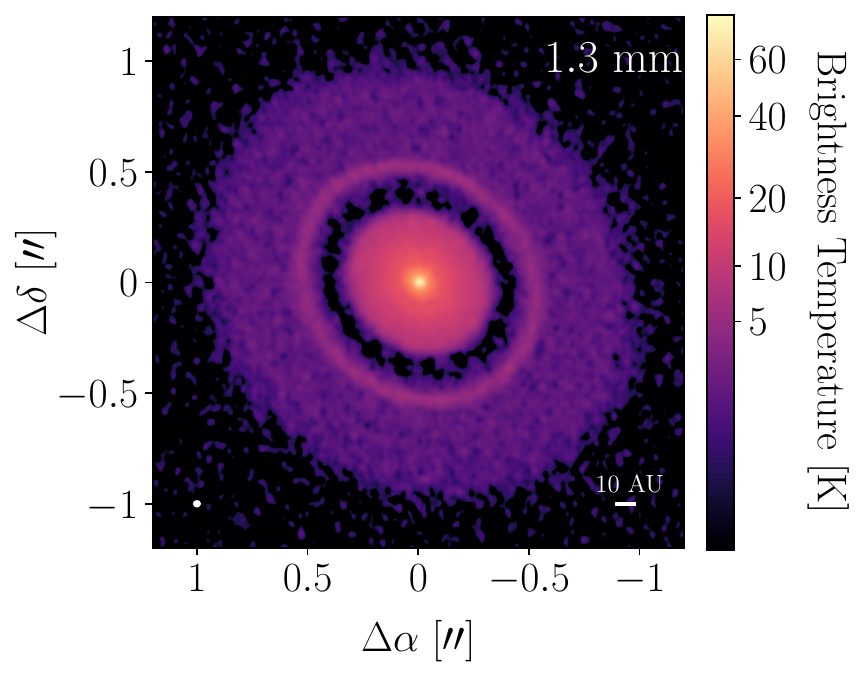}
    \includegraphics[width=0.49\linewidth]{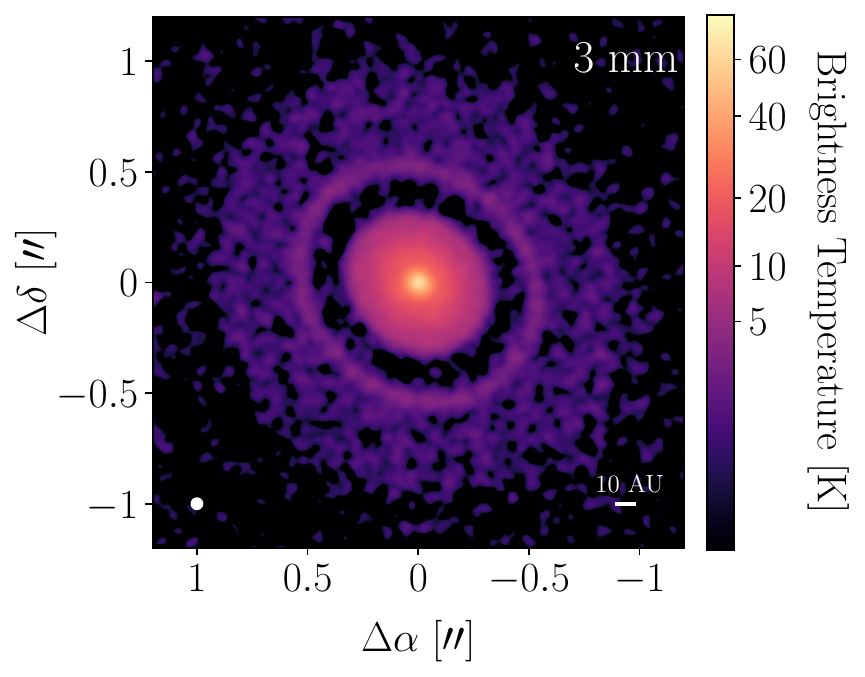}
     \caption{High resolution ALMA images of Elias\,24 at 1.3 mm and 3 mm, with an RMS of 0.35 K (19 $\mu$Jy beam$^{-1}$) at 1.3\,mm and 0.33 K (9 $\mu$Jy beam$^{-1}$) at 3\,mm. 
     The white ellipse in the bottom left shows the beam size of the image, corresponding to $0.036'' \times 0.034''$ (5.0 au $\times$ 4.7 au) at 1.3mm, $0.060'' \times 0.058''$ (8.2 au $\times$ 7.9 au) at 3mm. The white hash at the bottom right is 10 au. }
    \label{fig:CleanImages}
\end{figure*}

We present a multiwavelength (1.3 and 3\,mm), high angular resolution (29\,mas, 4.0\,au) study of the disk in Elias\,24.
These data allow us to directly study the impact of an embedded planet forming in a disk and theories of dust trapping due to that planet. 

This paper is structured as follows: in Section \ref{sec:data}, we describe the new 3\,mm data and its calibration and in Section \ref{sec:frank}, we compute deconvolved radial profiles at both wavelengths. From these profiles we infer dust properties in the disk and report our results in Section \ref{sec: dust properties}. A discussion of our results is presented in Section \ref{sec:discussion} and our conclusions in Section \ref{sec:conclusions}.

\section{Data and Calibration} \label{sec:data}
The 1.3 mm ALMA observations are from the Disk Substructures at High Angular Resolution Program \citep[DSHARP,][]{Andrews_DSHARP_I_2018ApJ}, and we employ their publicly available calibrated visibilities. The data span $uv$-distances of 11.5$-$11,500 k$\lambda$. The new 3 mm ALMA observations come from ALMA Programs 2017.1.01330.S (PI: Pérez, L) and 2018.1.01198.S (PI: Pérez, L), conducted on November 30, 2017, and June 12-29, 2019, with 3 executions in compact antenna configurations (C43-6, 15m$-$2,500m baselines) and 4 executions in extended antenna configurations (C43-10, 244m$-$16,200m), respectively. The resulting range of $uv$-distances in the 3 mm data are 5-5,400 k$\lambda$. 

The data were taken using ALMA Band 3 ($\lambda \approx 3$\,mm) in 4 spectral windows (SPWs) centered at 90.5, 92.5, 102.5, and 104.5 GHz. All SPWs had a 1.875 GHz bandwidth sampled at 128 channels each. The standard ALMA Pipeline Calibration was applied; for the 90.5 GHz SPW of the compact configuration data we find significant discrepancy in amplitude between the two independent correlations, so we omit this SPW and use only the three other SPWs for the short-baseline data. We image each execution to check the astrometry and phase-center of each observation (which may vary due to the astrometric uncertainty of ALMA and the proper motion of the source between observations). We then align the different executions to a common phase-center, determined by fitting a 2D Gaussian to the emission peak of one of the long-baseline observations. Finally, we note that offsets in flux amplitude between each execution were within uncertainties ($< 3 \%$) and required no correction.



Self-calibration was done with CASA versions 5.4.0 and 5.6.1 using the parallelized $\mathtt{mpicasa}$. We follow the  method described in \citet{Andrews_DSHARP_I_2018ApJ}  and \citet{Oberg_MAPS_2021ApJS}, using CASA 5.4.0 for tasks $\mathtt{gaincal}$ and $\mathtt{virtualconcat}$.
We self-calibrate the short-baseline observations first, starting with phase-only self-calibration, where we combined the SPWs and step down the solution intervals (scan-length, 90s, 45s, and 24s). We then performed one step of phase-amplitude self-calibration with a scan-length solution interval and combining all SPWs. This procedure resulted in a 220\% improvement in the SNR of the short-baseline data.
We combined the long-baseline data with the self-calibrated short-baselines, and together these were self-calibrated combining all SPWs and using 2 steps of phase-only self-calibration (solution interval of 300s and scan-length) and one step of phase-amplitude self-calibration, with an infinite solution interval. This procedure resulted in a 7\% improvement on the SNR of the combined dataset. 

We imaged the data with CASA $\mathtt{tclean}$, using the multiscale, multifrequency synthesis mode \citep{Cornwell_multiscale_2008ISTSP}, a Briggs weighting parameter of $\mathtt{robust}=0.8$\footnote{This value provided the best compromise between the final CLEAN image resolution and the resulting SNR.}, and a cleaning threshold of 1$\sigma$. We determined our 1$\sigma$ uncertainty for a given image by computed the RMS of the image in an annulus centered on the source and spanning 1.7 arcsec to 3.0 arcsec. For the 1.3 mm data we use the same CLEAN parameters as in \citet{Andrews_DSHARP_I_2018ApJ} and a 1$\sigma$ cleaning threshold. The final 3 mm CLEAN image reached a resolution of $0.060'' \times 0.058''$ (8.2 au $\times$ 7.9 au) and is shown in Figure \ref{fig:CleanImages}, alongside the 1.3 mm CLEAN image, which has a resolution of $0.036'' \times 0.034''$ (5.0 au $\times$ 4.7 au). We also produce a 3 mm image using a Briggs $\mathtt{robust}=0.0$, to obtain better angular resolution for the CLEAN-image-derived radial profiles discussed in Appendix \ref{app:Tests}.

Since we are combining two ALMA configurations for our data, we investigated the possible need for a JvM correction \citep{Czekala_MAPS_2021ApJS}. The correction factor for the 3 mm data is 0.98, so we do not apply any correction. For the 1.3 mm data, the correction factor is 0.78. However, the target is sufficiently bright at 1.3 mm that applying the correction to the image produced almost no change, besides to decrease the $RMS$ noise level by 20 \%. We also directly fit the visibilities rather than relying on the azimuthal averages of CLEAN images, so our radial profile results are not impacted by the mismatch between the true PSF and the CLEAN beam.

\section{Radial Profiles at 1.3 and 3 mm} \label{sec:frank}

To recover radial intensity profiles of higher spatial resolution than those extracted from $\mathtt{CLEAN}$ images via azimuthal averaging (i.e. to avoid beam convolution), we use the visibility fitting package $\mathtt{frankenstein}$ \citep[$\mathtt{frank}$ hereafter;][]{jennings_frank_2020MNRAS}. 
The code assumes an axi-symmetric emission distribution and fits a 1-D radial brightness profile to the real, deprojected visibilities using a Gaussian process. 
The \emph{deconvolved} radial intensity profiles of the Elias\,24 disk, obtained independently at 1.3 and 3\,mm, are presented in the top panel of Figure \ref{fig:frankProfiles}, while our fit  and uncertainties estimation are described in Appendix \ref{sec:franksims}.
These profiles have a finite intrinsic spatial resolution (treated here like the "beam" full width at half maximum, or FWHM) of $\theta_\mathrm{1.3 \ mm} = 15 \pm \ 0.1$ mas and $\theta_\mathrm{3 \ mm} = 29 \pm \ 0.1$ mas, at 1.3 and 3 mm, respectively, that we determine following the procedure outlined in Appendix \ref{sec:FrankResolution}.

The radial intensity profiles clearly show both the ring, B77, and the gap, D57, as identified in \citet{Huang_DSHARPII_2018ApJ}. B77 and D57, originally named for the physical stello-centric radii, are located at 0.564$^{\prime \prime}$ and 0.418$^{\prime \prime}$, respectively, corresponding to 78 au and 58 au using the new, 139 pc, Gaia distance to the system. However, we will adopt the historic naming of the features in this letter. Detailed analysis of the features is given in Section \ref{sec: dust properties}.

Since both profiles need to have the same resolution in order to compute a spectral index profile (with $I_\nu \propto \nu^{\alpha}$, where $\alpha$ is the spectral index of the emission), we convolve the 1.3 mm profile by $ \sqrt{\theta_\mathrm{3 \ GHz}^2 - \theta_\mathrm{1.3 \ mm}^2 } = 25$ mas to match the 3 mm profile resolution. We then compute the spectral index profile as $\alpha = \log(\frac{I_\mathrm{230 \ GHz}}{I_\mathrm{100 \ GHz}}) / \log(\frac{230}{100})$. The spectral index profile is shown in the bottom panel of Figure \ref{fig:frankProfiles}.

\begin{figure}[!htb]
    \centering
    \includegraphics[width=\linewidth]{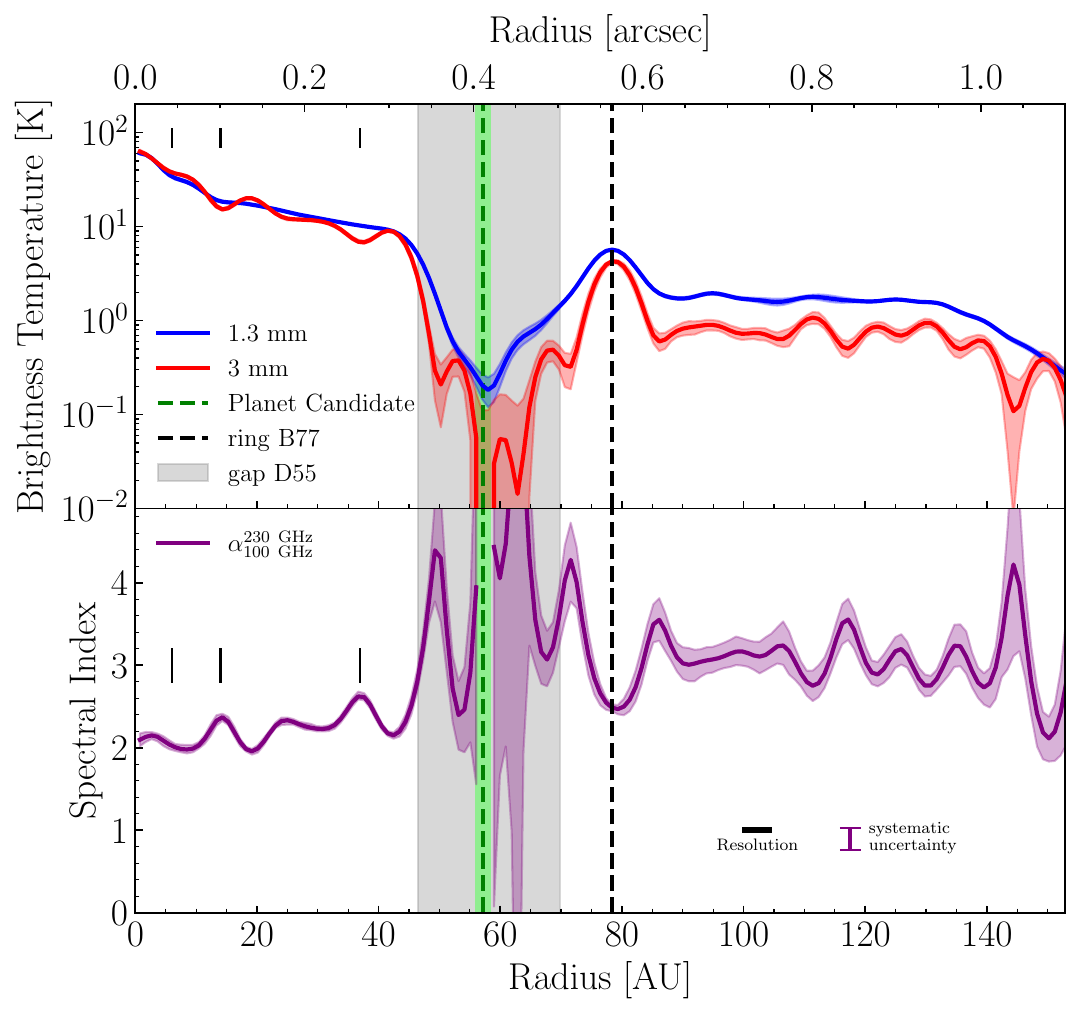}
    \caption{\textbf{Top:} Deconvolved radial intensity profiles computed from the 1.3 mm and 3 mm visibilities, with 1-$\sigma$ uncertainties shown as the blue and red shaded regions respectively. The green dashed line shows the semi-major axis of the protoplanet candidate reported in \citet{Jorquera_Elias24_CPD_2021AJ}, with the green shading showing the astrometric uncertainty in its position. The grey shading marks the gap location and the black dashed line marks the center of the ring, both as defined in \citet{Huang_DSHARPII_2018ApJ}. \textbf{Bottom:} The radial profile of the spectral index $\alpha$ computed from the profiles in the top panel. The black scalebar shows the resolution element of the spectral index profile, while the purple bar indicates the systematic uncertainty in the spectral index $\alpha$ due to ALMA's 10\% absolute flux-scale uncertainty. The vertical hashes mark the locations of the smaller gaps discussed in Section \ref{sec:substructure}.} 
    \label{fig:frankProfiles}
\end{figure}


\section{Dust and Ring Properties} \label{sec: dust properties}




\subsection{Dust-trapping efficiency of the ring} \label{sec:trapping}

First, we show that dust-trapping is taking place in the main ring, B77, following the procedure in \citet{Dullemond_DSHARP_2018ApJ} where the deconvolved radial width of the dust ring ($w_d$) is compared with the estimates of the gas ring pressure bump width. \citet{Dullemond_DSHARP_2018ApJ} show that in order to avoid being disrupted by instabilities like the Rossby wave instability \citep{Li_RossbyWaveInstability_2000ApJ}, the width of the gas ring should be larger than its pressure scale height ($h_p$). Therefore, if $w_d < h_p$, the dust ring is radially narrower than the gas ring can be, implying that some mechanism is trapping the dust particles radially. 

For simplicity, we assume that the underlying gas and dust ring radial profiles are Gaussian, an approximation that only breaks down near in the wings of the profiles \citep{Dullemond_DSHARP_2018ApJ}. We use the 1.3 mm $\mathtt{frank}$ fit at its native resolution of 15 mas (i.e. without convolving to the resolution of the 3 mm fit) to avoid artificially broadening the ring in the 1.3 mm prior to our measurement. 

To compute $w_d$, we fit a single Gaussian to B77 (within $78 \pm 5$ au), using the $\mathtt{frank}$ profile at 1.3 and 3\,mm. We report the Gaussian fit parameters and best-fit values in Table \ref{tab:gaussFits}. 
The ring widths (Gaussian standard deviation, $\sigma$) are $3.64 \pm 0.04$ au and $3.02 \pm 0.02$ au for the 1.3 mm and 3 mm profiles, respectively. The deconvolved dust widths are $w_d = \sqrt{\sigma^2 - \sigma_b^2}$ and the deconvolved amplitude is $A_\mathrm{dec} = \frac{\sigma}{w_d}A$, where $\sigma_b = \theta_{\mathrm{maj}}/(2\sqrt{2\ln{2}})$ is the beam FWHM ($\theta_{\mathrm{maj}}$) converted to a Gaussian standard deviation, and $A$ is the amplitude of the fitted Gaussian. The  $b_{\mathrm{maj}}$ values for the 1.3 mm and 3 mm profiles are the estimated $\mathtt{frank}$ resolutions, or $15 \pm 0.1$ mas and $29 \pm 0.1$ mas respectively. The calculated dust widths are 3.53 and 2.49 au for the 1.3 mm and 3 mm data, respectively. We report these and the deconvolved amplitudes in Table \ref{tab:gaussFits}. 

We estimate $h_p$ by adopting analytic prescriptions for the dust  pressure scale height and temperature profiles. For the dust temperature profile, we assume a passive disk and use the flared disk model from \citet{Chiang_Goldreich_passive_1997}, 
\begin{equation} \label{eq:tempProf}
    T_d(r) = \left( \frac{\frac{1}{2}\varphi L_*}{4 \pi r^2 \sigma_{SB}}  \right)^{\frac{1}{4}},
\end{equation}
where $r$ is the radial position in the disk, $\sigma_{SB}$ is the Stefan-Boltzmann constant, $L_*$ is the stellar luminosity, and $\varphi$ is the flaring angle of the disk. We use $L_* = 6 \ L_\odot$ \citep{Natta_El24Mass_2006} and assume $\varphi = 0.02$, as adopted in \citet{Dullemond_DSHARP_2018ApJ}. When we compare the passive disk temperature profile predictions with the brightness temperature of the disk for $r < 20$ au, the brightness temperature is 0.5$\times$ lower than predicted from the passive disk model. Since the emission is optically thick in this region and the brightness temperature should trace the actual dust temperature, we scale the profile by a factor of 0.5.


We then assume the gas and dust are in thermal equilibrium and compute the scale height assuming vertical hydrostatic equilibrium, which is given by 
\begin{equation}
    h_p(r) = \sqrt{\frac{k_B T_d(r) r^3}{\mu m_\mathrm{p} G M_*}},
\end{equation}
where $\mu=2.33$ is the mean molecular weight of the gas \citep[if dominated by H$_2$, as assumed by][]{Dullemond_DSHARP_2018ApJ}, $m_\mathrm{p}$ is the mass of a proton, $G$ is the gravitational constant, $k_B$ is the Boltzmann constant, and $M_*=0.78$ $M_\odot$ is the stellar mass \citep{Andrews_DSHARP_I_2018ApJ}. The ring peaks at the same radius of $r_0 = 78$ au at 1.3 mm and 3 mm, which we adopt as the ring center. Thus, we compute at the ring location a temperature of $T_d(78 \ \mathrm{au}) = 11$ K and a pressure scale-height of $h_p(78 \ \mathrm{au}) = 5.15$ au. We then find that $w_d < h_p$ at both wavelengths and that $w_d (3 \ \mathrm{mm}) < w_d (1.3) \ \mathrm{mm} $  ($2.49 \pm 0.02 \ \mathrm{au}$ versus $3.53 \pm 0.09 \ \mathrm{au}$). Both of these facts are strong indicators that radial dust-trapping is taking place at the main ring of Elias\,24, with the larger grains being more radially trapped than the smaller grains. 



\citet{Dullemond_DSHARP_2018ApJ} showed that assuming the radial settling of the dust grains can be related to the vertical settling, we can use the pressure scale height, $h_p$, and estimates of the maximum and minimum pressure bump widths, $w_\mathrm{max}$ and $w_\mathrm{min}$, respectively, to constrain the trapping efficiency of the ring. In general, the narrower the dust ring is, relative to the estimated pressure bump width, the more efficiently trapped the dust. 

The maximum pressure bump width, $w_\mathrm{max}$, is measured as the separation of the ring to the nearest minimum, so  $w_\mathrm{max}=17.1$ au. The minimum pressure bump width is $w_\mathrm{min} = h_p$ for the sake of stability in the ring. \citet{Dullemond_DSHARP_2018ApJ} then define a measure of the trapping efficiency in the bump as $\psi = \sqrt{\frac{\alpha_{\mathrm{turb}}}{\mathrm{Sc \ St}}}$, where $\alpha_{\mathrm{turb}}$ is the usual turbulence parameter, $\mathrm{Sc}$ is the Schmidt number (set to 1 for simplicity in our analysis\footnote{The Schmidt number gives the ratio between the turbulent viscosity and the turbulent diffusion coefficient. Since we are not aiming to constrain this, and previous work has found the ratio to be around 1.0 \citep[e.g.,][]{Johansen_dustDiffusion_2005ApJ}, this is a reasonable assumption.}) and $\mathrm{St}$ is the Stokes number. This trapping efficiency parameter can be constrained using the relationship between the maximum and minimum pressure bump widths described above and the measured dust ring width, $w_d$ according to:
\begin{equation}
    \psi_\mathrm{min} = \left[ \left( \frac{w_\mathrm{max}}{w_d} \right)^2  -1 \right]^{-\frac{1}{2}}, 
    \psi_\mathrm{max} = \left[ \left( \frac{w_\mathrm{min}}{w_d} \right)^2  -1 \right]^{-\frac{1}{2}}.
\end{equation}
From the estimated $\psi_\mathrm{min}$ and $\psi_\mathrm{max}$ values, we can compute $\frac{\alpha_\mathrm{turb}}{\mathrm{St}}$, which we report in Table \ref{tab:gaussFits}. We find that for both bands, the $\psi$ values are below 1.0 and therefore within the regime of efficient trapping described by \citet{Dullemond_DSHARP_2018ApJ}. 

Furthermore, our measurement of $w_d$ at 1.3 mm is identical (within uncertainty) to that reported by \citet{Jennings_frankDSHARP_2022MNRAS} in their $\mathtt{frank}$ analysis of the DSHARP data. With our 3 mm data we corroborate their finding that B77 is a strong candidate to be a potential dust trap.

\begin{deluxetable*}{ccccccccccccccc}[!htb]
	\tablecaption{Best-fit parameters of Gaussian function fits to the ring at 77 au.
 \label{tab:gaussFits}}
	\tablewidth{0pt}
	\tablehead{
	    \colhead{Band} & \colhead{A} & \colhead{$\sigma$} & \colhead{$r_0$}  & \colhead{$w_d$} & \colhead{$A_{\mathrm{dec}}$}  & \colhead{$\tau_\nu$\tablenotemark{*}} & \colhead{$w_{\mathrm{min}}$} & \colhead{$w_{\mathrm{max}}$} & \colhead{$\psi_{\mathrm{min}}$} & \colhead{$\psi_{\mathrm{max}}$} & 
     \colhead{$\left( \frac{\alpha_\mathrm{turb}}{\mathrm{St}} \right)_{\mathrm{min}}$} & \colhead{$\left( \frac{\alpha_\mathrm{turb}}{\mathrm{St}} \right)_{\mathrm{max}}$} \\
	    \colhead{} & \colhead{(Jy as$^{-2}$)} & \colhead{ (au) } & \colhead{(au)} & \colhead{(au)} & \colhead{(Jy as$^{-2}$)}  & \colhead{} & \colhead{(au)} & \colhead{(au)} & \colhead{} & \colhead{} & \colhead{} & \colhead{} & \colhead{}
	}
\startdata
    $1.3$ mm  &   $0.255 \pm 0.001$      &  $3.64 \pm 0.04$  &  $78.39 \pm 0.03$ & 3.53 & 0.262 & 5.47 & 5.15 & 17.1 & 0.21 & 0.91  & 0.04 & 0.84 \\ 
    $3$ mm  &     $0.0296 \pm 0.0001$   &  $3.02 \pm 0.02$   &  $78.74 \pm 0.02$ & 2.49 & 0.036  & 0.86 & 5.15 & 17.1 & 0.15 & 0.55 & 0.02 & 0.31  \\ 
\enddata
    \tablenotetext{*}{We compute the absorption-only optical depth using $\tau_\nu = -\mathrm{ln}\left( 1 - A_\mathrm{dec}/B_\nu(\mathrm{11 \ K})  \right)$}
\end{deluxetable*}

\subsection{Radiative transfer fits to $\Sigma_d$, $a_\mathrm{max}$, and $T_d$}
\begin{figure*}
    \centering
    \includegraphics[width=0.48\linewidth]{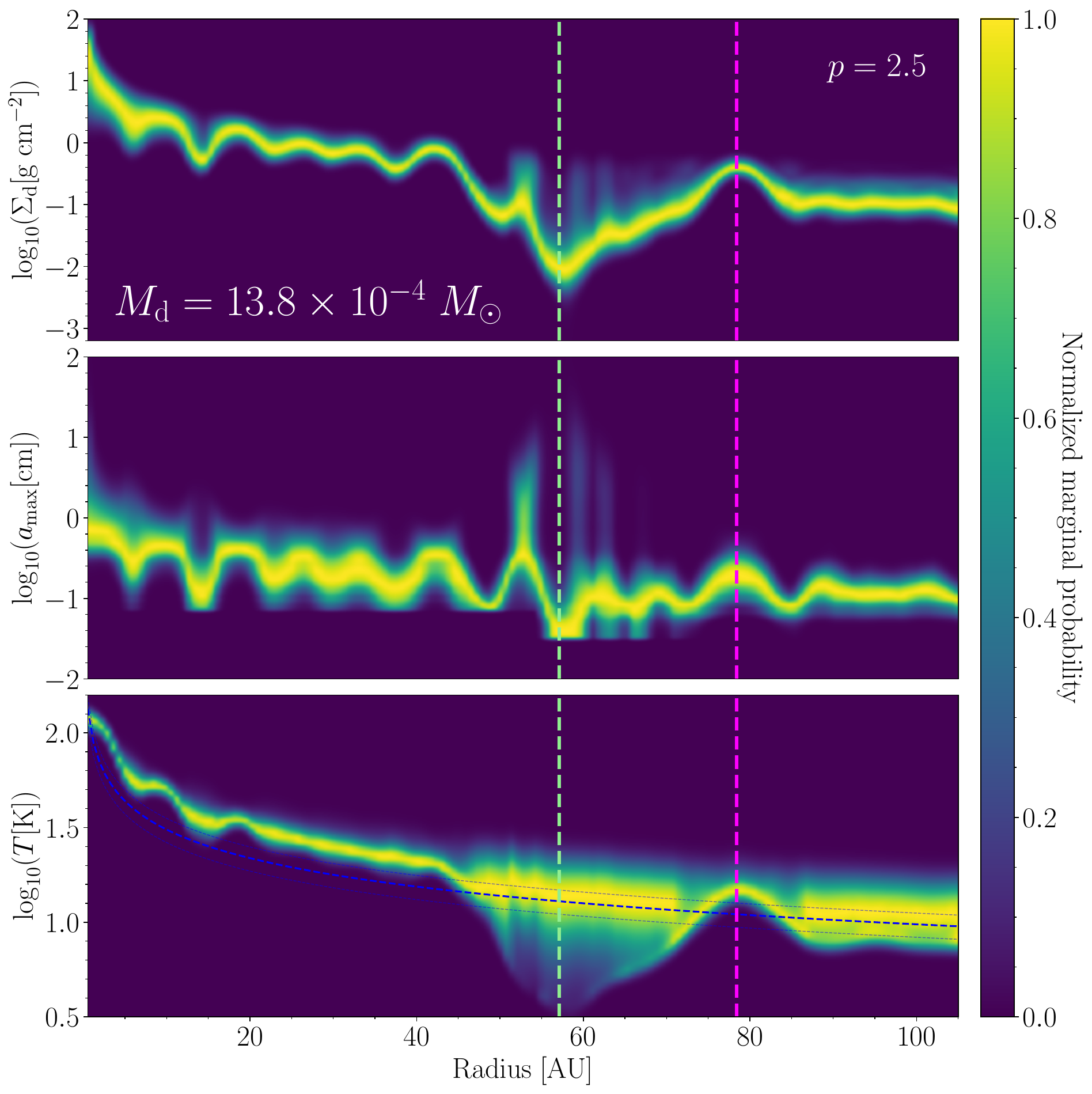}
    \includegraphics[width=0.48\linewidth]{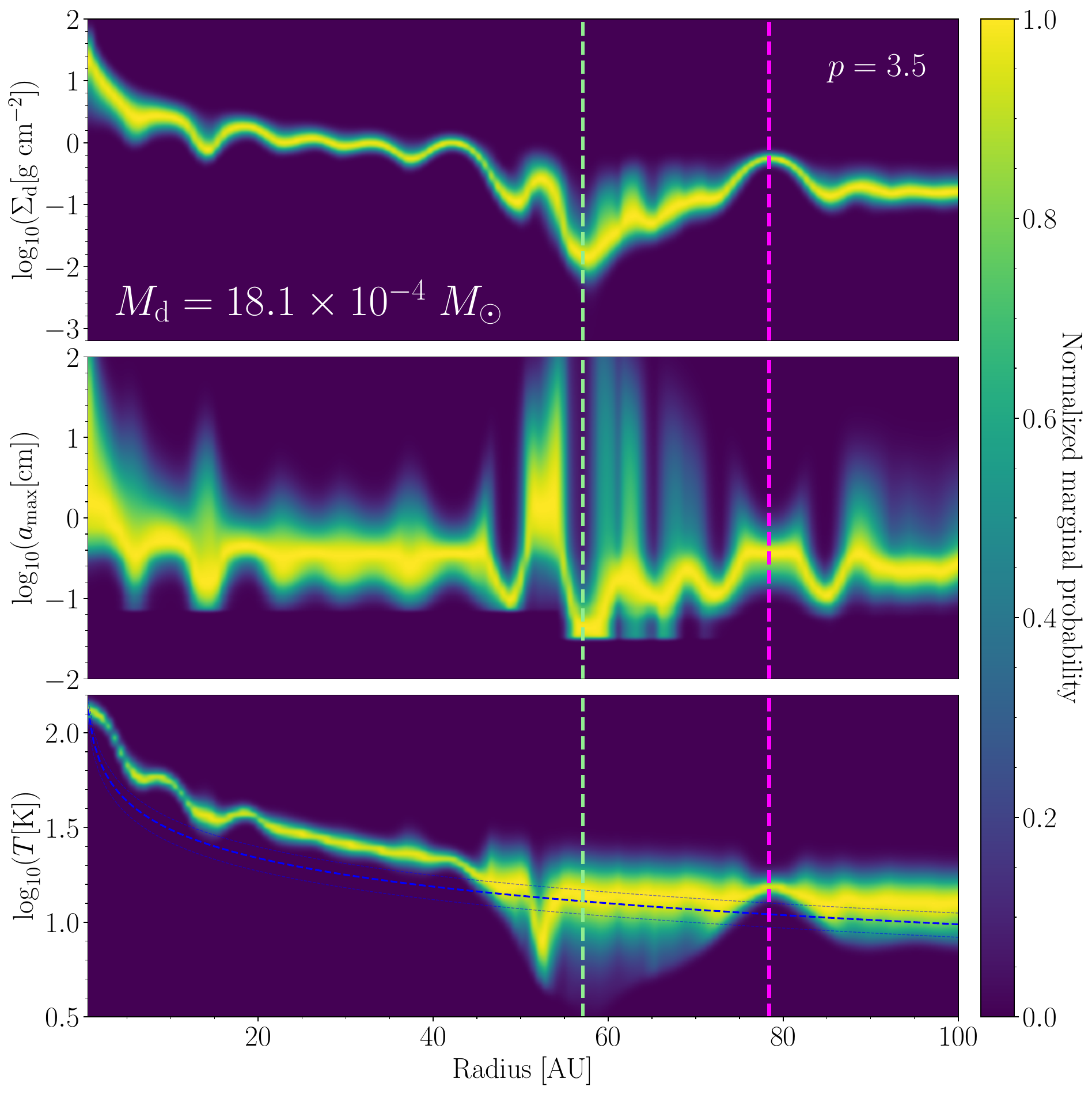}
    \caption{Radial profiles of the dust properties in the disk around Elias\,24. The left column shows our results for $p = 2.5$, while the right column shows $p = 3.5$. The orbital radius of the candidate companion from \citet{Jorquera_Elias24_CPD_2021AJ} is shown with a green dashed line and the ring location is shown with a magenta dashed line. \textbf{Top:} The dust surface density profile. Notice the increase in surface density at the ring location and the significant decrease in the gap. \textbf{Center:} The maximum grain size profile, which also increases at the ring location, indicating potential grain growth in the ring. \textbf{Bottom:} The temperature profile distributions. The blue dotted line shows the prior temperature profile given by Equation \ref{eq:tempProf}, with $1 \sigma$ Gaussian standard deviations shown as blue dashed lines.} 
    \label{fig:DustProfiles}
\end{figure*}

To determine the radial profiles of the dust surface density ($\Sigma_d$), maximum grain size ($a_\mathrm{max}$), and temperature ($T_d$), we perform radiative transfer modeling following the model in \citet{Sierra_MAPS_2021ApJS}, where we include the effect of dust scattering opacity and use the DSHARP dust opacity law\footnote{We have also tested using the \citet{Ricci_TaurusOpacities_GrainGrowth_2010A&A} opacities and find similar results for our $\Sigma_d$ and $T_d$ profiles, although the $a_\mathrm{max}$ values are on average 0.03 cm, or 10$\times$ smaller than when we use the DSHARP opacities.} \citep{birnstiel-DSHARP-scattering_2018ApJL}. However, we only have 2 bands with which to constrain the 4 model parameters ($\Sigma_d$, $T_d$, $a_\mathrm{max}$, and $p$, the dust distribution power law index) at each radial position. 
We therefore impose 2 additional constraints on the model: we fix the power law index of the grain size distribution to $p=2.5$ and impose a prior on the dust temperature profile. The choice of $p = 2.5$, which is shallower than the measured interstellar $p$ \citep{mathis-interstellar-grains-1977ApJ}, allows us to estimate a lower bound on the $a_\mathrm{max}$ values in the disk \citep[e.g., ][]{Sierra_MAPS_2021ApJS}. We also tested our models with $p = 3.5$ (right column of Figure \ref{fig:DustProfiles}) and found it does not significantly impact our posterior distributions, so we proceed with our $p=2.5$ models for the remainder of this paper. 

The prior we impose on $T_d(r)$ is that the probability of models with a temperature far from that given in Equation \ref{eq:tempProf} decreases following a Gaussian distribution. The standard deviation of the distribution is given by the 0.2 dex uncertainty on $L_*$ reported in \citet{Natta_El24Mass_2006}, which translates to an 11 \% uncertainty in $T_d(r)$, or $\pm 1.5$ K for the 11 K ring temperature. This helps with the degeneracy between $\Sigma_d$, $a_\mathrm{max}$, $T_d(r)$. 

We also impose restrictions on the maximum grain size parameter space. We note that one important assumption we make is requiring $a_\mathrm{max}(r < 50 \ \mathrm{au}) > 0.072$ cm. This is necessary break a degeneracy in our solutions between large and small grain sizes, particularly at $r < 20$ au. We justify this and other priors in detail in Appendix \ref{app:dustModelPriors}.

The normalized marginal probability distributions of the three parameters are shown in Figure \ref{fig:DustProfiles}. The posteriors are relatively well-confined, particularly for the $\Sigma_d$ and $T_d$ profiles. In general, the $T_d(r)$ profile favors slightly warmer values than the scaled passive disk profile we adopted as our prior and in Section \ref{sec:trapping}. For example, $T_d(78 \ \mathrm{au}) = 15$ K, rather than the 11 K we adopted to estimate the pressure scale height, which would give $h_p \sim 6$ au. The change would result in a slightly greater trapping efficiency than what we calculate in Table \ref{tab:gaussFits}. 

We note that at $r = 55-60$ au, the 3 mm profiles have several $I_\nu = 0$ values, which produce an artifact in the posterior probability distributions at that location. Inside the gap, the posteriors are quite broad, but favor slightly values of $\Sigma_d$ and $a_\mathrm{max}$ interior to and exterior to the orbit of the proposed planet, while plummeting at the proposed orbit radius. 

At the ring location, both the $a_\mathrm{max}$ and $\Sigma_d$ are localized maxima, a sign that larger dust grains are piling up in the ring.
This implies the ring is acting as a dust trap, as predicted in the case of a planet clearing a gap \citep[e.g., ][]{pinilla_ring_shaped_dust_2012A&A, Nazari_multiband_models_2019MNRAS}. 

Integrating our $\Sigma_d(r)$ profile gives a total dust disk mass of $M_\mathrm{dust} = 13.8^{+0.7}_{-0.5} \times 10^{-4} \ M_\odot$ for the p = 2.5 models. If we assume the standard $\Sigma_g/\Sigma_d = 100$, we find $M_\mathrm{disk} = 0.138 M_\odot$, indicating a particularly massive disk. However, in the following sections we discuss the likelihood of a much lower gas-to-dust ratio in this disk. The $\Sigma_d$ estimate may also potentially be lowered by assuming some degree of porosity in the dust grains \citep{Zhang_porosity_2023ApJ}. However, without polarimetric data the degree of grain porosity is difficult to constrain. 



\subsection{The Toomre Q and Stokes Number Profiles} \label{sec:ToomreQ}
From the modeling described in Section \ref{sec: dust properties} we compute the distributions of other important quantities: the Stokes number and the Toomre $Q$ parameter (discussed in this Section), and the absorption-only and scattering + absorption optical depths (shown in Appendix \ref{app:OpticalDepths}). 

To compute the Toomre $Q$ parameter, we combine the probability distributions of $\Sigma_d$ and $T(r)$ by taking $10^6$ random samples from both at each radius value. We then use the following expression:
\begin{equation} \label{eq:Q}
    Q(r) = \frac{c_s(r) \Omega(r)}{\pi G \Sigma_g(r)},
\end{equation}
where $\Omega(r) = \sqrt{G M_*/r^3}$ is the Keplerian angular velocity, $c_s(r)= \sqrt{k_B T(r)/2.34 m_H}$ is the local sound speed, $\Sigma_g(r) = 100 \ \Sigma_d(r)$ is the local gas surface density, $k_B$ is the Boltzmann constant, $m_H$ is the mass of a hydrogen atom, and $G$ is the universal gravitational constant. 

We follow a similar procedure to compute the Stokes number, $\mathrm{St}$, using the expression
\begin{equation} \label{eq:St}
    \mathrm{St}(r) = \frac{\pi \rho_m a_\mathrm{max}(r)}{2 \Sigma_g(r)},
\end{equation}
where $\rho_m=1.675$ g cm$^{-3}$ is the average grain density \citep[][]{birnstiel-DSHARP-scattering_2018ApJL} and $a_\mathrm{max}(r)$ is the maximum grain size.

\begin{figure*}[!htb]
    \centering
    \includegraphics[width=0.46\linewidth]{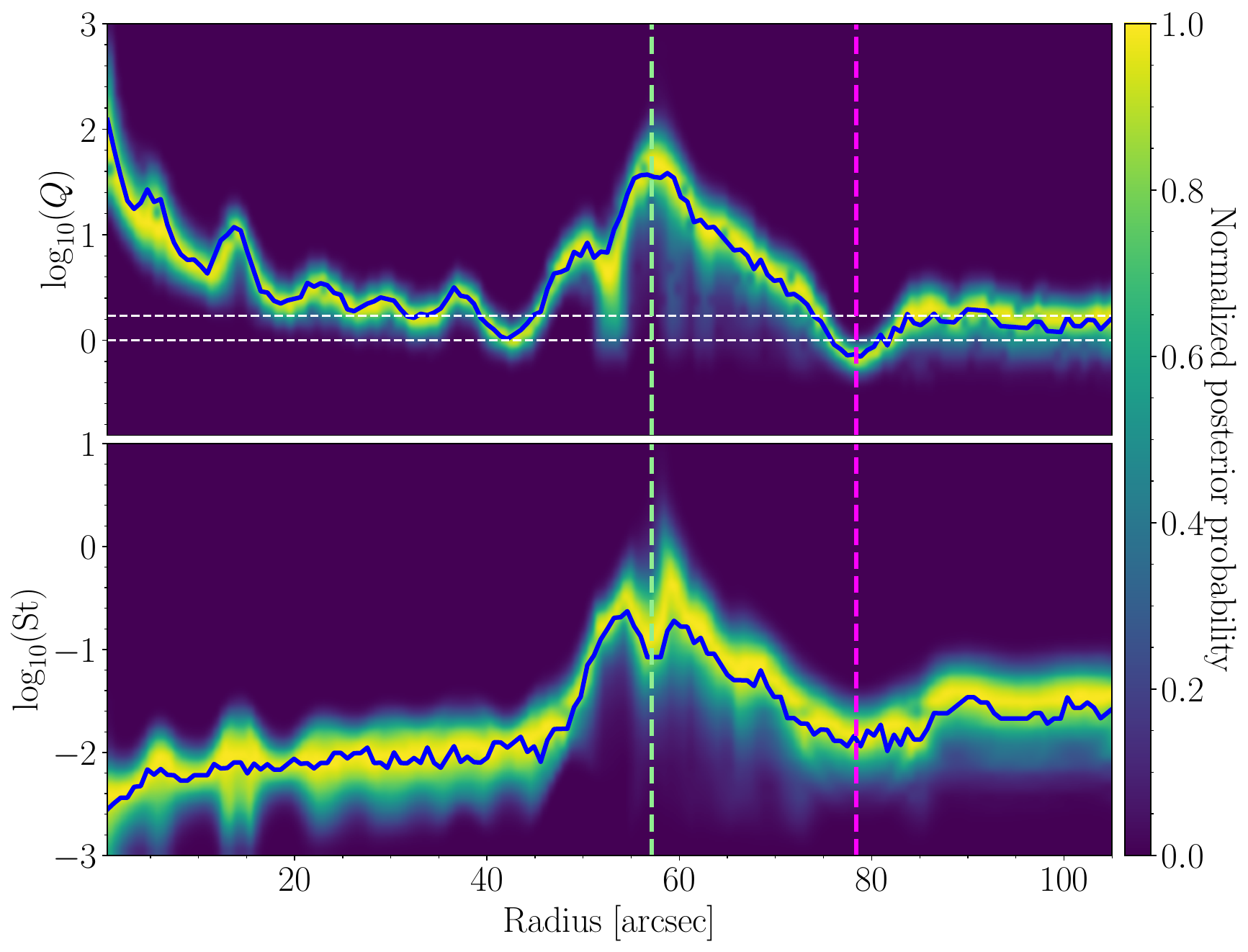}
    \includegraphics[width=0.50\linewidth]{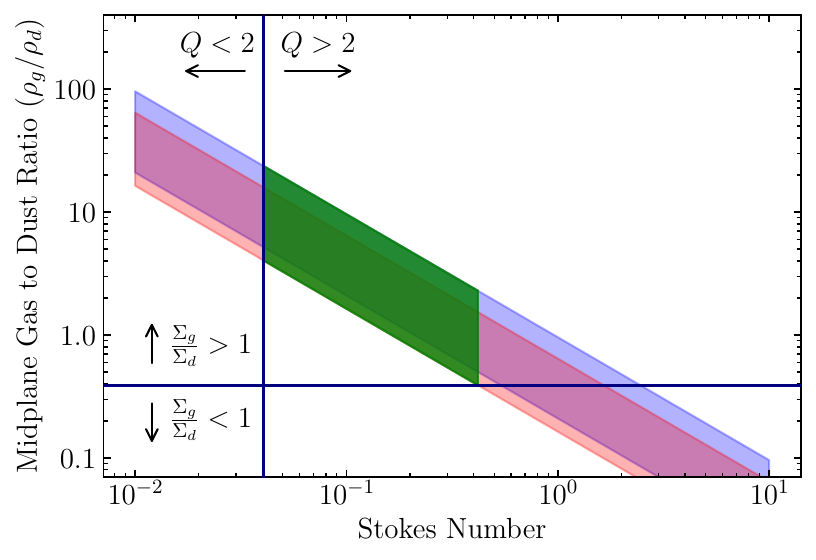}
    \caption{\textbf{Left:} The marginal probability distributions of the Toomre $Q$ profile (upper panel) and the Stokes number profile (lower panel) sampled from the distributions shown in Figure \ref{fig:DustProfiles}. The candidate companion orbit radius is shown with the green dashed line and the ring center is shown with the magenta dashed line. The solid blue line marks the values preferred by the overall best-fit model. The white dashed lines in the upper panel show the $Q = 1$ and $Q = 1.7$ lines. Assuming a gas-to-dust ratio of 100, there are many locations in the disk where $Q < 1.7$, implying the disk would be gravitationally unstable. \textbf{Right:} The range of midplane gas to dust ratios in B77 estimated in Section \ref{sec:GasToDust}. The blue and red shaded regions indicate the upper and lower limits from the maximum and minimum $\sqrt{\alpha_\mathrm{turb}/\mathrm{St}}$ values in Table \ref{tab:gaussFits} for the 1.3 mm (blue) and 3 mm (red) data. The lower limit on allowed Stokes number (vertical dark blue line) is obtained by requiring gravitational stability, since we do not see spirals in the disk. The lower limit on $\rho_g/\rho_d$ (horizontal dark blue line) is obtained by requiring the $\Sigma_g > \Sigma_d$. The "allowed" values between these limits are marked by the green shaded region.     }    
    \label{fig:Toomre}
\end{figure*}

The $Q$ and $\mathrm{St}$ distributions are shown in Figure \ref{fig:Toomre}. There are several locations in the ring that would be gravitationally unstable ($Q < 1.7$) if the gas-to-dust ratio really were 100 as is assumed in our profiles. However, in that case, the disk would be expected to have strong spirals, which we do not see in our continuum images (Figure \ref{fig:CleanImages}) or in the radial-profile-subtracted residuals \citep[Appendix \ref{app:residuals} and][]{Andrews_CPD_2021ApJ}. Nonetheless, the St and Q profiles in Figure \ref{fig:Toomre} can give some intuition for how both constants vary in the disk. In Section \ref{sec:GasToDust}, we estimate the expected $\Sigma_g/\Sigma_d$ ratio in the ring and demonstrate it should be $<$ 50 to ensure stability. 

The optical depth profiles including scattering in addition to absorption are presented in Appendix \ref{app:OpticalDepths}. They indicate that outside of the gap the disk is nearly optically thick at both 1.3 and 3 mm ($\tau \sim 0.5$), and is very optically thick in the inner disk ($\tau > 2$). The optical depth at the location of the ring, $\tau \sim 1.0$, is greater than in the rest of the outer disk. 


\subsection{Constraints on the Gas-to-Dust Ratio in B77} \label{sec:GasToDust}
The Toomre Q profiles shown in Figure \ref{fig:Toomre} show that if we assume a gas-to-dust ratio of 100 everywhere, the disk should be gravitationally unstable in many locations, including in the ring. However, we do not see any of expected observational signatures of gravitational instability in the disk or ring. Therefore, the gas-to-dust ratio should be much lower than 100. Constraining the ratio directly would require a measurement of the gas mass in the disk, which is difficult for Elias 24 due to absorption from the envelope \citep{Andrews_DSHARP_I_2018ApJ}. 

From the modeling described in Section \ref{sec: dust properties} and the constraints on the dust dynamics in B77, derived in Section \ref{sec:trapping}, we can constrain the gas-to-dust ratio in B77. We can do this by relating the optical depth in B77 to the Stokes number via the DSHARP opacities, as is done in \citet{Stammler_DustGasRatio_2019ApJ}. 

We begin by writing the absorption-only optical depth as $\tau_\nu = \kappa_\nu \Sigma_d$, where $\kappa_\nu$ is the dust opacity. Expressing $\kappa_\nu$ in terms of a dimensionless quantity $\xi_\nu$ gives us
\begin{equation} \label{eq:opac_q}
    \tau_\nu = \kappa_\nu \Sigma_d = \frac{\pi a_\mathrm{max}^2}{m}\xi_\nu \Sigma_d = \frac{3}{4}\frac{\xi_\nu}{a_\mathrm{max} \rho_m} \Sigma_d,
\end{equation}
where $\rho_m=1.675$ g cm$^{-3}$ is the average grain density \citep[][]{birnstiel-DSHARP-scattering_2018ApJL}. We can then combine Equations \ref{eq:opac_q} and \ref{eq:St} to get
\begin{equation}
    \frac{\Sigma_g}{\Sigma_d} = \frac{3 \pi}{8} \frac{\xi_\nu}{\tau_\nu} \frac{1}{\mathrm{St}}.
\end{equation}
Since we have $\Sigma_d = 0.38$ g cm$^{-2}$, $a_\mathrm{max} = 0.21$ cm, and $\tau_\nu$ in the ring (see Appendix \ref{app:OpticalDepths} for our $\tau_\nu(r)$ profiles), we can compute $\xi_\nu / \tau_\nu$ directly. So our final step is to convert the surface densities to midplane volume densities $\rho_d$ and $\rho_g$: $\rho_d = \Sigma_d / (\sqrt{2\pi} h_d)$ and $\rho_g = \Sigma_g / (\sqrt{2\pi} h_p)$, where $h_d$ is the dust scale height. The dust scale height can be related to the gas scale height by $h_d \sim h_p \sqrt{\alpha_\mathrm{turb}/ (\alpha + \mathrm{St})}$ \citep{Dubrulle_dust_gas_scaleheights_1995Icar}, which is the ratio we estimated for bands in Section \ref{sec:trapping}. So our final expression becomes:
\begin{equation}
    \frac{\rho_g}{\rho_d} = \frac{3 \pi}{8} \frac{\xi_\nu}{\tau_\nu} \sqrt{\frac{\alpha}{\alpha + \mathrm{St}}} \frac{1}{\mathrm{St}}.    
\end{equation}

Sampling several values of St, and using the maximum and minimum $\alpha_\mathrm{turb}/\mathrm{St}$ ratios in Table \ref{tab:gaussFits}, we get the range of expected $\rho_g/\rho_d$ values shown in Figure \ref{fig:Toomre}. We constrain the lowest allowable Stokes number by requiring that the disk be gavitationally stable ($Q > 2$). Although this criterion is sensitive to our assumed temperature profile, we note that increasing the assumed dust temperature in the disk to 26 K results in only a 10\% increase in $\rho_g/\rho_d$ for all St values. We also place a lower bound on the gas to dust ratio by requiring $\Sigma_g/\Sigma_d > 1$. 

With our gravitational stability constraint, the maximum $\rho_g/\rho_d$ value allowed in the ring is 25, corresponding to a $\Sigma_g/\Sigma_g = 27$, much lower than the 100 we assume in Section \ref{sec:ToomreQ}. \citet{Stammler_DustGasRatio_2019ApJ} demonstrated that the large $\tau_\nu \sim 1$ optical depth in B77 is indicative that the environment in the ring may be favorable to triggering streaming instability. In Figure \ref{fig:Toomre}, the range of $\rho_g/\rho_d$ values for St $> 0.04$ approach the critical $\rho_g/\rho_d \sim 1$ that would be most conducive to the streaming instability \citep{YoudinGoodman_streamingInstability_2005ApJ}. Adopting the numerical-simulation-derived constraints based on $\Sigma_g/\Sigma_d$ and St in \citet{Lesur_HydroDynamicsOfPPDs_PPVII_2023ASPC}, B77 is in the ``strong clumping" regime and highly conducive to the streaming instability for St $> 0.04$.

\section{Discussion} \label{sec:discussion}
We begin our discussion by identifying the substructure that can be seen at both 1.3 mm and 3 mm and how the features differ as a function of wavelength. We then shift the focus of our discussion to the two dominant features in the Elias\,24 radial profiles: the bright ring at 78 au, which we conclude is a dust trap, and the deep gap centered at 58 au, which shows excess emission near the gap edges. The high angular resolution in the data, boosted by the visibility fitting in $\mathtt{frank}$, enables us to study the morphology of the gap-ring pair in detail.

\subsection{Substructure observed in both bands} \label{sec:substructure}
The radial profiles in both bands are very similar, with slightly more substructure observable in the 3 mm radial profile than in the 1.3 mm profile. This may be due to the relatively higher signal-to-noise of the 3 mm data, especially in the inner disk, or could indicate that the larger grains trace disk substructure more sensitively. 

Starting close to the center of the disk, at $\sim 6$\,au, we observe a small drop in intensity -- potentially a marginally resolved gap. Next is the gap at 14 au reported for the 1.3 mm data in \citet{Jennings_frankDSHARP_2022MNRAS} and \citet{Andrews_CPD_2021ApJ}. This gap is deeper at 3 mm, indicating the feature may indeed be a previously unresolved gap in the system. The gap is followed by a small ring, seen again at both 1.3 mm and 3 mm, coincident with a drop of the spectral index in the lower panel of Figure \ref{fig:frankProfiles}. Both gaps are also seen in the 1.3 mm $\mathtt{frank}$ profile shown in Appendix \ref{app:Tests}, but are lost in the convolution to the 3 mm profile resolution.

Unfortunately, the inner disk substructure is not clearly seen the $a_\mathrm{max}$ profile.  In the \citet{Guidi_HD163296_2022A&A} analysis of HD 163296, the high optical depth at $r < 40$ au produces large uncertainties in their maximum grain size estimates. We encounter a similar problem for $r < 30$ au in Elias\,24. Despite constraining the temperature profile tightly, we still see $\sim 0.5$ dex uncertainties in the $a_\mathrm{max}$ profile in this region.

At 35-40 au, we see another slight decrease in the 3 mm flux, a potential third gap, followed by a bump at 40-45 au, preceding the 58 au gap. Neither feature is appears in the 1.3 mm profile. The CLEAN image radial profiles shown in Appendix \ref{app:Tests} also have a small gap at $r \sim 38$ au followed by a bump in the 3 mm but not in the 1.3 mm data. A small bump preceding a gap is consistent with hydrodynamical simulations for a giant planet carving a gap in a disk like the Elias\,24 disk \citep{Nazari_multiband_models_2019MNRAS}. 

The large gap-ring pair in the system, centered at 58 au and 78 au, respectively, is clearly seen at both 1.3 and 3\,mm. Significant previously unresolved structure is present in the gap at 3 mm and also at 1.3 mm. The gap has a clear minimum at 58 au, coincident with the radius of the point source reported in \citet{Jorquera_Elias24_CPD_2021AJ}. In this gap minimum, the emission is a factor of $0.003$ times the peak flux in the disk. Surrounding the minimum, there is "shoulder"-like emission in both bands, which is $10 \times$ brighter than the gap minimum. The emission may be the result of dust remaining in the gap as a planet first starts to clear out material; an intermediate stage between a fully cleared gap and a uniform disk. We discuss this further in Section \ref{sec:gap}. The ring at 78 au is noticeably narrower at 3 mm than at 1.3 mm. We discuss the difference in width and physical reasons for this in Section \ref{sec: dust properties}.
The diffuse emission beyond 80 au does not show much significant substructure at 1.3 mm. The radial profile is mostly flat at larger radii. The "wiggles" seen in the 3 mm data are likely due to the low signal-to-noise in the outermost region of the disk. In Appendix \ref{app:Tests} we show that the wiggles also appear in the radial profiles computed by azimuthally averaging the CLEAN images.

\subsection{The ring (B77)} \label{sec:ring}

We have presented in Section \ref{sec: dust properties} strong evidence of dust trapping and grain growth in the Elias\,24 main ring. Following the procedure described by \citet{Dullemond_DSHARP_2018ApJ} to estimate the trapping properties of a ring, we see that it presents efficient trapping in both 1.3 and 3 mm (though slightly more efficient in 3 mm). The first immediate piece of evidence for this is that the ring is much more narrow at 3 mm than it is at 1.3 mm. We also estimate low $\psi$ values, in both bands, which indicates the dust is well-settled and narrowly confined in the pressure bump. The relatively low $\frac{\alpha_\mathrm{turb}}{\mathrm{St}}$ values we calculate ($\sim 0.02-0.84$) correspond to an $\alpha_\mathrm{turb}$ range of $8\times 10^{-4}$ to $0.03$ (assuming a minimum Stokes number of $0.04$ discussed in Section \ref{sec:GasToDust}). While a wide range, this does favor the intermediate/lower values found previously in other works for other disks \citep{pinte_HLTau_settling_2016ApJ, villenave_ALMAedgeOnDisks_2020A&A, villenave_settledDisk_2022ApJ}. 

Modeling the dust emission following a method similar to that described in \citet{Sierra_MAPS_2021ApJS}, we find stronger evidence of trapping. Both the dust surface density and maximum grain size are significantly greater at $r \sim 78$ au than for $r > 80$ au. While $a_\mathrm{max}(r > 80$ au$) \sim 0.16$ cm, $a_\mathrm{max}(r = 78)$ au $ = 0.63$ cm, which is similar to the $a_\mathrm{max}$ values in the inner disk ($r < 40$ au). In the surface density profile we see $\Sigma_d(r = 78)$ au$= 0.38$ g cm$^{-2}$, whereas $\Sigma_d (r > 80$ au)$\sim 0.08$ g cm$^{-2}$, a factor of 4.8 enhancement in the dust surface density. This is evidence that not only is there a pileup of dust in the ring, but the dust grains may actually be conglomerating and growing. Dust grain growth in the ring would explain the large increase in the maximum grain size at that location. The dearth of large grains in the outer disk, seen in both our radial profiles and more directly in the weak 3 mm emission at $r > 80$ au, is circumstantial evidence of the rapid migration of larger grains. 

The spectral index behavior in the ring and gap is also consistent with that reported in previous multiband analyses of disks like HL Tau \citep{ALMA_HLTau_2015ApJ}, TW Hya \citep{Huang_TwHya_Multiband_2018ApJ, Macias_TWHya_2021A&A}, GM Aur \citep{Huang_GMAur_Multiband_2020ApJ}, and HD 163296 \citep{Guidi_HD163296_2022A&A, Doi_HD163296_Multiband_2023ApJ}. In those targets, the spectral index sharply increases at gap locations and rapidly decreases again at ring locations. For rings at smaller radii, the spectral index approaches a Rayleigh-Jeans value of 2, as we see in Elias\,24. Fitting the emission from TW Hya, \citet{Macias_TWHya_2021A&A} also find a significant increase in dust surface density and maximum grain size at their ring locations. 

We also find that, assuming the canonical $\Sigma_g/\Sigma_d = 100$, the ring reaches $Q < 1$, which would be highly unstable and susceptible to fragmentation (see Figure \ref{fig:Toomre}, left panel). Even for $r > 80$ au, the disk is only marginally stable ($Q \sim 1.7$) under this assumption. If we impose the constraint that the ring should be gravitationally stable, we estimate a maximum gas-to-dust ratio of 27. Several other regions of the disk have $Q < 1.7$ in Figure \ref{fig:Toomre}, implying that the gas-to-dust ratio should be lower ($\sim 50$ for many of these regions) to ensure the disk is gravitationally stable.

Using a similar argument, \citep{Ohashi_DGTau_SurfDens_2023ApJ} show that at $\Sigma_g/\Sigma_d < 10$ in the DG Tau disk in order to maintain stability, particularly for $10 \ \mathrm{au} < r < 30 \ \mathrm{au}$. However, their angular resolution is relatively low ($> 100$ mas) and may not entirely rule out signatures of gravitational instability on smaller scales. Multiband analyses of HL Tau and CW Tau \citep{carrasco-gonzalez_HLTauDustDist_2019ApJ, Ueda_multiband_CWTau_2022ApJ} find that both disks are stable when assuming $\Sigma_g/\Sigma_d = 100$ despite having similar $\Sigma_d(r)$ profiles to Elais 24 due to the higher $T(r)$ in HL Tau ($T(70 \ \mathrm{au}) \sim 45$ K) and the much more compact disk in CW Tau.


The Stokes number in B77 is $>0.04$ according to our stability argument, which may be higher still if trapping decreases the gas to dust ratio (see the right panel of Figure \ref{fig:Toomre}). Even this minimum Stokes number is sufficiently high to encourage planetesimal growth via the streaming instability \citep{Schaffer_streamingInstability_2021A&A, Lesur_HydroDynamicsOfPPDs_PPVII_2023ASPC}.



\subsection{The gap (D55)} \label{sec:gap}

Both radial profiles show significant structure in the gap, and much of it rises above the approximate $1 \sigma$ noise level of the radial profiles (see Section \ref{sec:frank}). Just interior and exterior to the gap minimum, there is a flat emission feature in both bands, much less steep than the gap walls. This may be consistent with material not fully cleared out yet, as the gap appears relatively shallow compared with the much steeper gaps seen in the AS\,209 and HD\,143006 disks \citep{Andrews_CPD_2021ApJ, Jennings_frankDSHARP_2022MNRAS}.

Approaching the gap minimum, the emission steeply drops, especially at 3 mm. Between $r = 42$ au and $r= 52$ au, the brightness temperature in both bands decreases by a factor of 25, which corresponds to a factor of 30 decrease in the surface density. The gap is extremely deep between $r = 55$ au and $r= 63$ au, where at approximately the orbital separation of the point source reported in \citet{Jorquera_Elias24_CPD_2021AJ}, $\Sigma_d(58 \ \mathrm{au}) = \Sigma_d(42 \ \mathrm{au})/250$. The optical depth also decreases dramatically in the gap, reaching a minimum of $\tau_\mathrm{1.3 \ mm} = 0.1$ and $\tau_\mathrm{3 \ mm} = 0.01$ (see Appendix \ref{app:OpticalDepths}).

We find that the gap/ring morphology in our radial intensity and spectral index profiles is similar to hydrodynamical simulations of gaps carved by massive planets, including those of \citet{Zhang_DSHARPVII_GapCarving_2018ApJ}, \citet{Nazari_multiband_models_2019MNRAS}, and \citet{Binkert_GapClearingPlanets_2021MNRAS}. The width, depth, and ``shoulder'' emission seen in D55 qualitatively match the features in 0.3-1 $M_\mathrm{Jup}$ and $\alpha_\mathrm{turb} \sim 10^{-3}$ models in these works, which are consistent with the mass limits for the system proposed in \citet{Jorquera_Elias24_CPD_2021AJ} and our estimated constraints on $\alpha_\mathrm{turb}$. 

Unfortunately, there are few detailed studies showing predictions of multiwavelength observations of planet-carved gaps in the Elias 24 system, so we focus our discussion to comparison with \citet{Nazari_multiband_models_2019MNRAS}. They simulate a migrating 30 $M_\earth$ planet in a system similar to Elias\,24 and in their Intermediate or Fast migration models, the spectral index rises quickly upon entering the gap, then flattens or decreases near the location of the protoplanet, then sharply decreases again upon reaching the ring. We see these features in our high-resolution spectral index profile as well. 


In all of their migration cases, however, \citet{Nazari_multiband_models_2019MNRAS} find the gap minimum slightly offset from the location of the migrating planet, and they predict a dust excess in the orbit of the protoplanet. Both of these are inconsistent with our radial profiles, though it is possible that the observations do not probe the predicted excess because it would likely be confined in azimuth, such as at Lagrangian points \citep[as seen in the LkCa 15 disk, ][]{Long_LkCa15_LagrangePoints_2022ApJ}. In the case of the non-migrating planets simulated by \citet{Zhang_DSHARPVII_GapCarving_2018ApJ}, the co-orbital excess emission is weaker for the lower mass planets, potentially consistent with a more strict upper limit on the mass of the Elias 24 companion. In the case of the \citet{Binkert_GapClearingPlanets_2021MNRAS} simulations, however, the excess emission co-orbital with the planet is up to 10$\times$ brighter than the gap minimum, even for the lower mass models, which is totally inconsistent with our observations. Critically, their model assumed an inviscid disk, which would enable more co-orbital dust to accumulate than in models with $\alpha \neq 0$ like we see in Elias 24. Further simulations of young planets carving gaps may be needed to better understand the evolution of any dust excess in the gap over time. 

\section{Conclusions} \label{sec:conclusions}

Our analysis of the 1.3 and 3 mm observations of the Elias\,24 disk using $\mathtt{frank}$ allow us to derive radial intensity profiles at high spatial resolution (3.9 au), which show consistent substructure in both bands. We find clear evidence of efficient dust trapping in  B77, the bright ring in the Elias\,24 system. Our analysis shows an increase in the local dust surface density and maximum grain size, as well as a localized minimum in the spectral index between the two bands. We also find that the dust is well-confined by the estimated pressure bump at the ring location, another sign that the ring is a dust trap. Elias 24 is the first system where a candidate planetary-mass companion has been directly detected inside a deep gap and where the localized kinematic perturbation that such an embedded planet produces has been detected. The fact that the B77 ring is trapping solids strengthens the planetary-mass companion hypothesis. 

Our multiband fits to the radial intensity profiles allow us to estimate the radially distributed dust properties in the Elias\,24 disk, from 10 au to 140 au. From our dust surface density profile, we are able to estimate a total solid mass of $M_\mathrm{dust} = 13.8^{+0.7}_{-0.5} \times 10^{-4} \ M_\odot$. Converting this to a gas mass using the standard ISM conversion ($M_\mathrm{gas} \sim 100 \ M_\mathrm{dust}$), we get $M_\mathrm{gas} = 0.138 \ M_\odot$. However, since we do not observe signatures of gravitational instability in the disk (see Section \ref{sec:ToomreQ}), the gas-to-dust ratio is likely much lower, possibly $< 50$, implying a maximum $M_\mathrm{disk} \sim 0.07 \ M_\odot$. 

We also find excess dust emission around the edges of the gap, consistent with an excess of large grains just interior to and some exterior to the proposed orbit of a candidate companion. Dust excess at radii near the radius of a forming planet may constrain the gap-clearing efficiency of the planet and help determine the formation time of the object. We recommend further study on the evolution of planet-cleared gaps over time, to test the expected gap geometries at different points in time once a Jupiter-mass planet begins clearing a gap. 

\section{Acknowledgements}
This work was supported in part by the NRAO Student Observing Support award SOSPADA-008. L.P. and A.C. gratefully acknowledge support from ANID FONDECYT Regular \#1221442.
L.P. gratefully acknowledges support by the ANID BASAL project FB210003. A.S. acknowledges support from FONDECYT de Postdoctorado 2022 \#3220495. L.P., A.S., M.B., acknowledge support from Programa de Cooperación Científica ECOS-ANID ECOS200049.

This project has received funding from the European Research Council (ERC) under the European Union’s Horizon 2020 research and innovation programme (PROTOPLANETS, grant agreement No. 101002188)

T.B. acknowledges funding from the European Union under the European Union's Horizon Europe Research and Innovation Programme 101124282 (EARLYBIRD) and funding by the Deutsche Forschungsgemeinschaft (DFG, German Research Foundation) under grant 325594231, and Germany's Excellence Strategy - EXC-2094 - 390783311. Views and opinions expressed are, however, those of the authors only and do not necessarily reflect those of the European Union or the European Research Council. Neither the European Union nor the granting authority can be held responsible for them.

This paper makes use of the following ALMA data: ADS/JAO.ALMA\#2013.1.00498.S, ADS/JAO.ALMA\#2017.1.01330.S, and \\ ADS/JAO.ALMA\#2018.1.01198.S. ALMA is a partnership of ESO (representing its member states), NSF (USA) and NINS (Japan), together with NRC (Canada), MOST and ASIAA (Taiwan), and KASI (Republic of Korea), in cooperation with the Republic of Chile. The Joint ALMA Observatory is operated by ESO, AUI/NRAO and NAOJ. 

The National Radio Astronomy Observatory is a facility of the National Science Foundation operated under cooperative agreement by Associated Universities, Inc.

\bibliography{Adolfo_bib}{}
\bibliographystyle{aasjournal}

\appendix 

\restartappendixnumbering

\section{$\mathtt{frank}$ fitting} \label{sec:franksims}

In Section \ref{sec:frank}, we show the deconvolved 1-D radial intensity profiles derived from visibility fitting. Here, we give the details of the $\mathtt{fit}$ and the Monte-Carlo technique we use to estimate the uncertainty on each profile. 

In general, $\mathtt{frank}$ fits 1-D radial brightness profiles to the observed (deprojected) visibilities using a Gaussian process \citep{jennings_frank_2020MNRAS}. The hyper-parameters of the fit are: $R_{\mathrm{max}}$, $N$, $p_0$, $\alpha$, and $w_{\mathrm{smooth}}$. $R_{\mathrm{max}}$ is the radius beyond which $\mathtt{frank}$ assumes emission to be 0. For Elias\,24, we use $R_{\mathrm{max}} = 2.1^{\prime \prime}$ (292 au), which is approximately twice the radius where the disk emission exceeds the RMS noise of the image. $N$ defines the number of grid points on which to sample the radial profile, which we set to 300 (giving a resolution element of 2.1$^{\prime\prime}/300 = 0.007^{\prime\prime}$. The $p_0$ parameter regularizes the power spectrum of the emission, and we set that to the recommended value of $10^{-15}$ Jy$^2$ \citep[see][]{jennings_frank_2020MNRAS}. The remaining hyperparameters, $\alpha$ and $w_{\mathrm{smooth}}$,  determine the lowest acceptable signal-to-noise ratio and act as prior weights for regularization, respectively. We discuss these further below. In our fits we take the non-negative solution offered by $\mathtt{frank}$, wherein the flux is never allowed to be $< 0$.

There is some degeneracy between $\alpha$ and $w_{\mathrm{smooth}}$ on any $\mathtt{frank}$ fit. For our fiducial radial profiles in Section \ref{sec:frank}, we set a conservative $\alpha=1.3$ and $w_{\mathrm{smooth}}=10^{-1}$, following \citet{Andrews_CPD_2021ApJ}. However, the choice of $\alpha$ and $w_{\mathrm{smooth}}$ is non-trivial and can alter the best-fit radial profile. Allowing lower $\alpha$ values frees the fit to use more of the (lower SNR) long-baseline visibilities, producing a higher-resolution profile. However, if $\alpha$ is set too low ($< 1.05$ in the case of 1.3\,mm data), it introduces high (spatial) frequency artifacts into the profile. Changes to the disk geometry (phase-center, inclination, and position angle) alter the deprojected visibilities and therefore also result in different profiles.

As is reported in \citet{jennings_frank_2020MNRAS}, we find that the formal uncertainties reported by $\mathtt{frank}$ are significantly underestimated, in light of the sources of error described above. To account for the uncertainties in the radial profile produced by different choices of $\alpha$, $w_{\mathrm{smooth}}$, and disk geometry, we compute 5000 different $\mathtt{frank}$ fits for the 1.3 mm and 3 mm data, randomly sampling hyperparameters and disk geometries. The priors for the hyperparameters and disk geometries are given in Table \ref{tab:FrankPriors}, and are either uniform, normal, or an asymmetric normal distribution described by,
\begin{equation}
    AN(x,  \sigma_1, \bar{x},\sigma_2) = 
    \left\{
        \begin{array}{lr}
            N(x, \bar{x}, \sigma_1) , & \text{if } x \leq \bar{x} \\
            N(x, \bar{x}, \sigma_2), & \text{if } x > \bar{x}
        \end{array}
    \right.
\end{equation}
These choices of prior distributions reflect the knowledge derived in fitting the disk geometry at 1.3mm by \citet{Huang_DSHARPII_2018ApJ}, while for the 3 mm data we fit the disk geometry using MCMC \citep[via the $\mathtt{emcee}$ package in Python,][]{FM_emcee_2013PASP} to minimize the imaginary component of the deprojected visibilities and find values consistent with the \citet{Huang_DSHARPII_2018ApJ} values, so we adopt those for consistency with the 1.3 mm data. 
To reduce the computational time of each fit, we perform the averaging described in \citet{Jennings_frankDSHARP_2022MNRAS} -- averaging our spectral windows to 1 channel each and binning in 30 s time intervals.

For Band 3 and Band 6, we construct Gaussian priors using the the inclination and position angle values and $1\sigma$ uncertainties from \citet{Huang_DSHARPII_2018ApJ}. We construct the phase-center prior distributions for Band 6 from the \citet{Huang_DSHARPII_2018ApJ} values. For the Band 3 phase-center prior distributions, we use the values from the MCMC posteriors.

The longest baselines of both datasets do not converge to 0, a sign that there is unresolved emission at the phase-center, which can introduce strong high-frequency oscillations in the $\mathtt{frank}$ fits. To correct for this, we subtract the emission of a point-source at the phase center given by 0.953 mJy at 1.3mm and 0.47 mJy at 3mm. The former is the mean of the longest baselines reported in \citet{Jennings_frankDSHARP_2022MNRAS}, and the latter was computed as the weighted mean of the visibilities greater than 3.5 M$\lambda$.

We compute our fiducial $\mathtt{frank}$ fits using the the full visibility tables (i.e., not using any channel and time averaging) with $\alpha=1.3$ and $w_{\mathrm{smooth}}=10^{-1}$ for both bands. We then turn to our 5000 $\mathtt{frank}$ fits and find the 16 and 84th percentile intensity profiles. We adopt these as our lower and upper 1 $\sigma$ uncertainties on the fiducial, non-averaged profiles.
The resulting profiles are shown in Figure \ref{fig:frankProfiles}, with the uncertainties displayed as shaded regions. The small variation between these 5000 fits demonstrates that the choice of different geometry and $\mathtt{frank}$ parameters does not dramatically impact the inferred radial profiles. We recover several features in the 1.3 mm data reported in \citet{Huang_DSHARPII_2018ApJ}, \citet{Andrews_CPD_2021ApJ}, and \citet{Jennings_frankDSHARP_2022MNRAS}. We also see these features in the 3 mm data, and discuss them in more detail in Section \ref{sec:frank}. 



\begin{deluxetable}{cccc}[!htb]
	\tablecaption{Prior distributions of parameters and geometries used in the $\mathtt{frank}$ simulations. $U[a,b]$ denotes a uniform distribution sampled from $a$ to $b$. $N(\bar{x}, \sigma_x)$ denotes a normal distribution with mean $\bar{x}$ and a standard deviation $\sigma_x$. $AN(\sigma_1, \bar{x}, \sigma_x)$ denotes an asymmetric normal distribution centered on $\bar{x}$ and with a standard deviation of $\sigma_1$ for $x < \bar{x}$ and $\sigma_2$ for $x > \bar{x}$. The construction of this asymmetric normal distribution is described in Appendix \ref{sec:franksims}.   \label{tab:FrankPriors}}
	\tablewidth{0pt}
	\tablehead{
	    \colhead{Parameter} & \colhead{units} & \colhead{1.3 mm} & \colhead{3 mm} 
	}
    \startdata
    $\alpha$  & &                $U[1.05, 1.3]$ &    $U[1.05, 1.3]$  \\ 
    {$\log_{10}$} $w_{\mathrm{smooth}}$  & &                $U[-3, -1]$ &    $U[-3, -1]$ \\ 
    $\Delta x$ & (mas) &                $N(-110.8, 0.8)$ &   $AN(0.08, -3, 0.19)$     \\ 
    $\Delta y$ & (mas) &                $N(-386.8, 0.9)$ &    $AN(0.09, 0.2, 0.12)$    \\ 
    $i$   & (deg) &                $N(29.0, 0.4)$ &       $N(29.0, 1.0)$ \\ 
    PA  & (deg) &                $N(45.7, 0.8)$ &        $N(45.7, 1.0)$ \\ 
    \enddata
\end{deluxetable}

\section{Determining the resolution of a $\mathtt{frank}$ profile} \label{sec:FrankResolution}
To be able to directly compare the 1.3 mm and 3 mm radial profiles, they should be convolved to the same angular resolution. However, the angular resolution of a given $\mathtt{frank}$ profile is not readily apparent. We compute the intrinsic angular resolutions of the radial profiles by sampling a ``delta'' ring ($\delta(r - r_0)$ at $r_0 = R_{\mathrm{max}}/2$ \footnote{$R_\mathrm{max}/2$ is chosen to represent a feature in the middle of the $\mathtt{frank}$ profile fit.}) at the same $(u,v)$ positions of the actual observations. We then fit those visibilities using the same procedure as before, thus, measuring the broadening of the resulting ring that $\mathtt{frank}$ is able to recover. The visibilities of the ring are computed by the Hankel transform of the delta ring as:
\begin{equation}
    \mathcal{H} \left[ \frac{A_0}{2 \pi r_0} \delta(r - r_0) \right] = A_0 J_0 (2 \pi r_0 q)
\end{equation}
where $r$ is the radial disk coordinate, $J_0$ is the zeroth order Bessel function of the first kind, $q = \sqrt{u^2 + v^2}$ is the deprojected baseline computed from the wavelength normalized baselines $u, v$, and $A_0$ is a constant representing the total disk flux. We 
fit a Gaussian profile to the intensity profile produced by $\mathtt{frank}$, and take only the non-negative solution, for methodological consistency with our data fits. 

We find the 1.3 mm $\mathtt{frank}$ radial profiles have an intrinsic resolution (Gaussian FWHM) of 15 $\pm \ 0.1$ mas, whereas the 3 mm profiles have an intrinsic resolution of 29 $\pm \ 0.1$ mas. Assuming the difference in resolution is given by $\theta_\mathrm{diff} = \sqrt{\theta_\mathrm{3 \ mm}^2 - \theta_\mathrm{1.3 \ mm}^2 }$, we convolve the 1.3 mm data by $\theta_\mathrm{diff} = 25$ mas 
, to bring it to the 29 mas resolution of the 3 mm data. This allows us to compare the profiles directly and compute the spectral index profile. The spectral index is computed as $I_\nu \propto \nu^{\alpha}$, where $\alpha$ is the spectral index of the emission. The spectral index profile is shown in the bottom panel of Figure \ref{fig:frankProfiles}. 

Though we do not analyze the excess emission features in the gap extensively, it is of note that the emission detection is of high significance. To determine this significance level conservatively, we estimate the $1 \ \sigma$ noise levels of the radial profiles relative to the CLEAN image noise. This can be done at a given radial position in the profile by dividing the $RMS$ noise of the CLEAN image by the square root of the number of beams contained in that radial bin. In the gap location ($r = $ 0.42$^{\prime \prime}$), we find that there are $\sim 45$ resolution elements. The $RMS$ noise levels in the CLEAN images, in units of $10^{10}$ Jy sr$^{-1}$, are 0.036 and 0.016 in the 1.3 mm and 3 mm data respectively. Using the 29 mas beam size and the 45 beams in the annulus centered on the gap, we get an improved rms of $1.36 \times 10^{-2}$ and $6.1 \times 10^{-3}$. For both datasets, then, the 1 $\sigma$ noise level in the radial profile is much lower than the features we identify in the gap.


\section{Additional Priors Imposed on the Radiative Transfer Dust Models} \label{app:dustModelPriors}

Due to degeneracy between solution sets dominated by small grains and those dominated by large grains \citep{Sierra_MAPS_2021ApJS}, we found it necessary to impose priors on the dust grain distribution radial profiles shown in Section \ref{sec: dust properties}. The priors were physically motivated to help produce more sensible fits. For instance, we see the large grain/small grain degeneracy distinctly in the $r < 20$ au region of the disk as a region of bimodal solutions, favoring either $a_\mathrm{max} \sim 1$ cm or $a_\mathrm{max} \sim 0.01$ cm. The bimodality in solutions appears suddenly and is not attributable to any physical feature we observe in the disk. If the small grain solution were selected it would produce a discontinuous jump by 2 orders of magnitude in the maximum grain size radial profile. Since we have no reason to expect a rapid transition in maximum grain size and we expect radial drift to produce a larger population of dust grains in the innermost region of the disk, we opt for a grain size distribution that favors the larger grain solution. We require that $a_\mathrm{max} (r < 50 \ \mathrm{au}) > 0.072$ cm, which we identify as the approximate boundary between the two families of solutions. 

We also find that there are regions in the inner disk where solutions with unrealistically high $\Sigma_d$ values ($\sim 100$ g cm$^{-2}$) are favored when there is no prior on the $\Sigma_d$ distribution. To guide the models away from extreme $\Sigma_d$, we also impose a power-law prior inspired by the stellar initial mass function (IMF) \citep[$dN/dM_* \propto M_*^{-2.3}$, ][]{Kroupa_IMF_review2001} on the expected dust surface density at each location. We do this by computing the ratio of $\Sigma_d$ at a radius $r$ to the expected solid mass of the minimum mass solar nebula at that location, $\Sigma_{d, \odot} = 20 \left( \frac{r}{1 \ \mathrm{au}} \right)^{-1}$. We compute the relative probability of this ratio by the IMF power law given above, $\left( \Sigma_d(r) / \Sigma_{d, \odot}(r) \right)^{-2.3}$, and then add it to the $\chi^2$ probabilities. The effect of this is to discourage the most extreme values of $\Sigma_d$ by nudging the model toward lower mass solutions. Our final $\Sigma_d(r)$ posterior distribution is well-constrained and produces a reasonable (though still large) $M_d$.

\section{The Optical Depth Profiles} \label{app:OpticalDepths}

\begin{figure}[!htb]
    \centering
    \includegraphics[width=\linewidth]{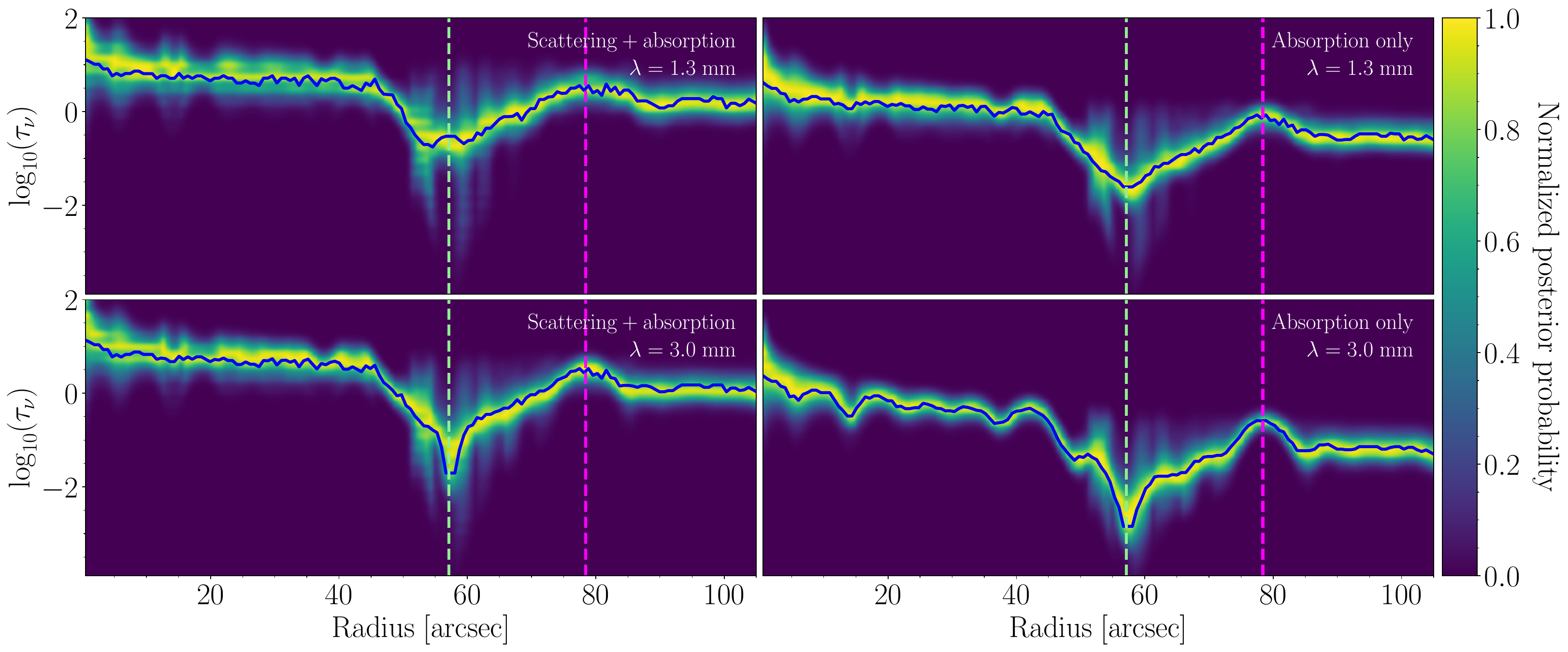}
    \caption{The scattering + absorption (left column) and absorption only (right column) optical depth marginal probability distributions. The candidate companion from \citet{Jorquera_Elias24_CPD_2021AJ} orbit radius is shown with the green dashed line and the ring location is shown with the magenta dashed line. The solid blue line shows the prefered values of the best-fit model. \textbf{Upper panels:} the optical depth profiles for the 1.3 mm data. \textbf{Lower panels:} the optical depth profiles for the 3 mm data. Notice that, when accounting for scattering, the "bumpy" features beyond 90 au disappear.}
    \label{fig:OpticalDepths}
\end{figure}

Here, we present the absorption only and scattering + absorption optical depth profiles (both shown in Figure \ref{fig:OpticalDepths}) computed using the procedure described in \citet{Sierra-scattering-2020ApJ}.

\section{Tests of the emission detected in the Frank profiles} \label{app:Tests}

\begin{figure}[!htb]
    \centering
    \includegraphics[width=0.49\linewidth]{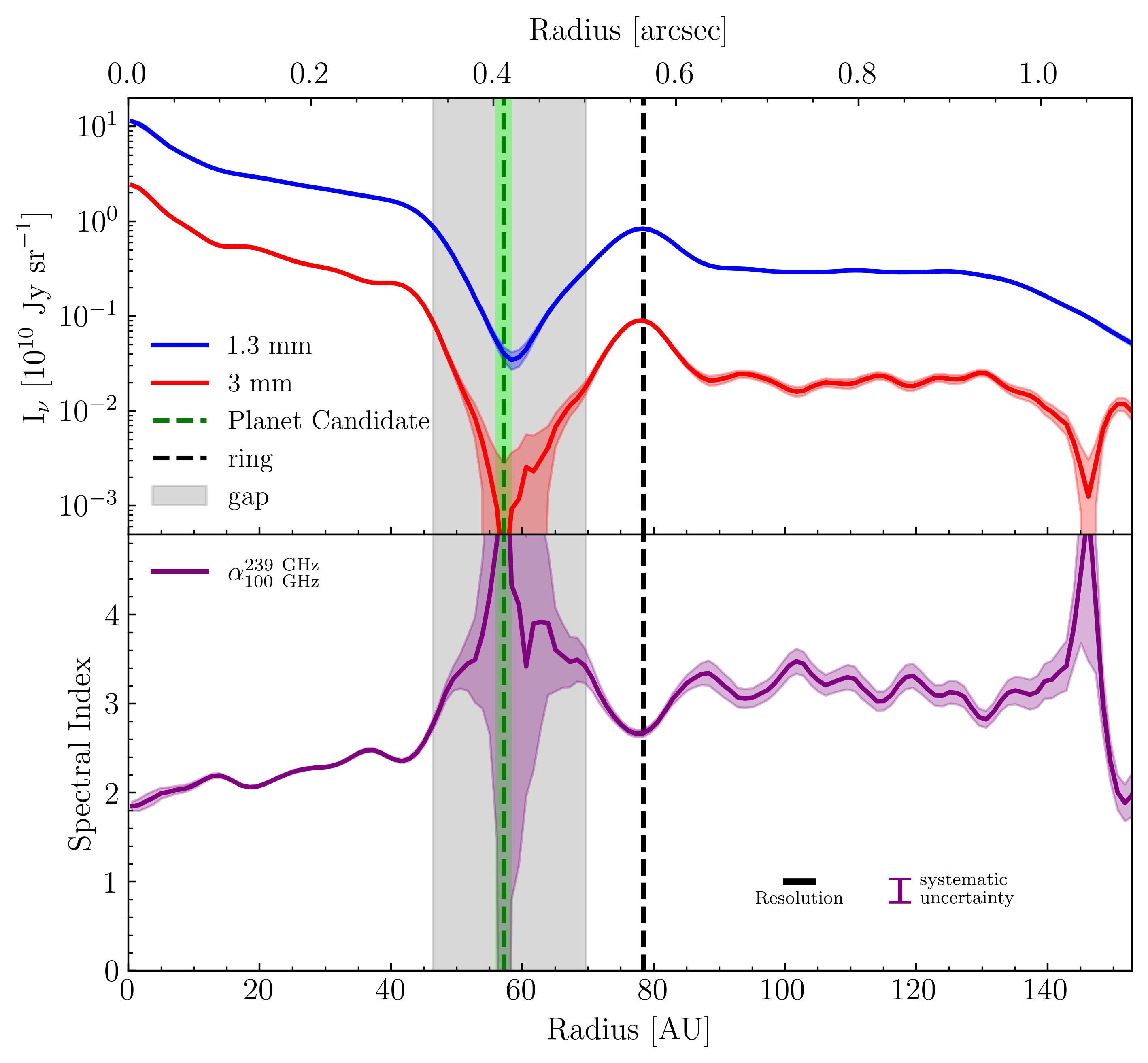}
    \includegraphics[width=0.49\linewidth]{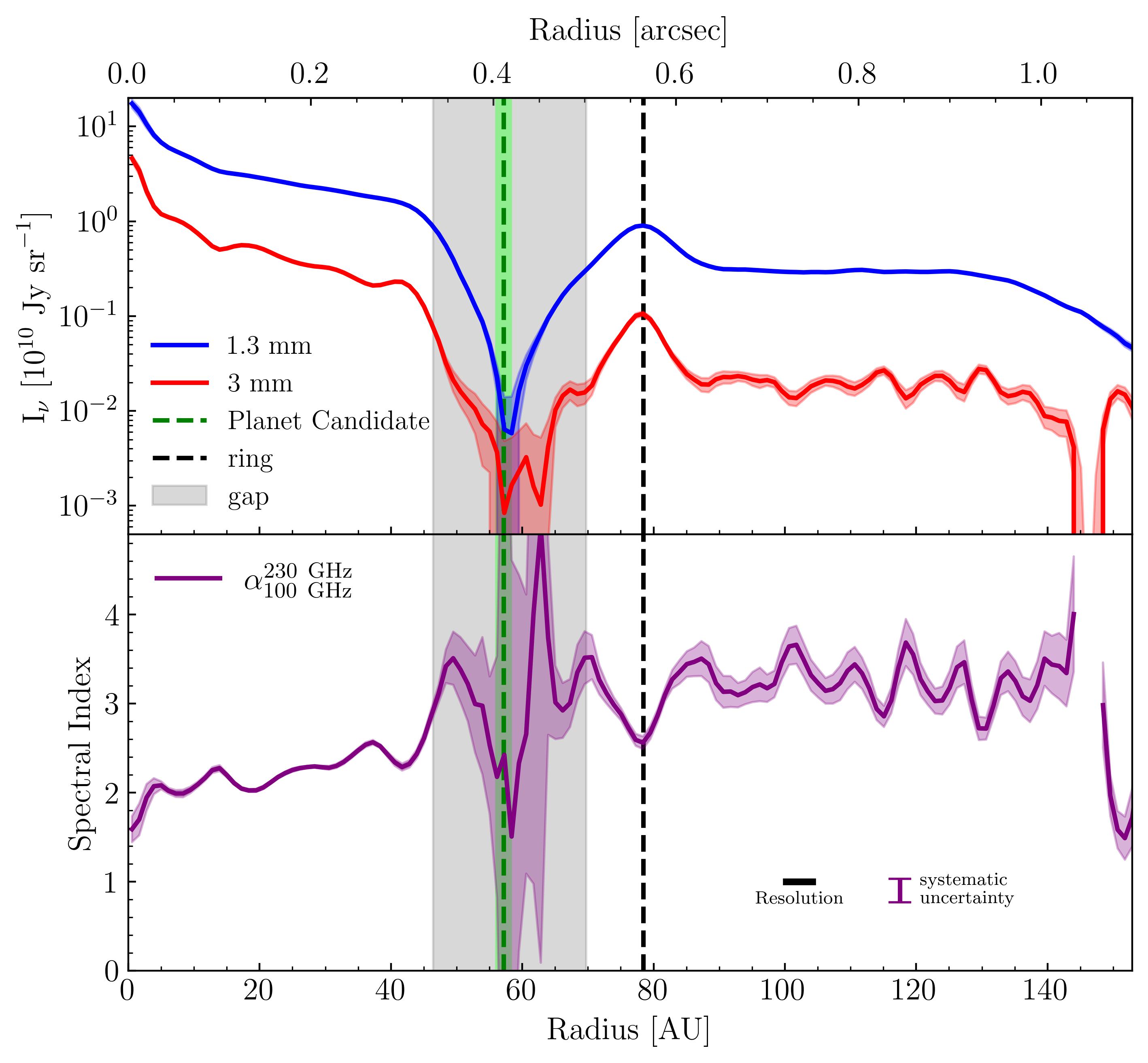}
    \caption{\textbf{Left Upper:} $\mathtt{CLEAN}$ image azimuthal average profiles computed from the 1.3 mm and 3 mm images, with the scatter from the binning shown as the blue and red shaded regions respectively. \textbf{Left Lower:} The radial profile of the spectral index computed from the $\mathtt{CLEAN}$ image profiles in the top panel. The black Gaussian profile shows the size of the resolution element of the profile, while the purple bar shows the systematic uncertainty due to the absolute flux uncertainty of ALMA. \textbf{Right Upper:} $\mathtt{CLEAN}$ model azimuthal average profiles computed from the 1.3 mm and 3 mm models, with the scatter from the binning shown as the blue and red shaded regions respectively. \textbf{Right Lower:} The radial profile of the spectral index computed from the $\mathtt{CLEAN}$ model profiles in the top panel. }
    \label{fig:frankProfiles_CLEAN}
\end{figure}

To confirm the appearance of features in our $\mathtt{frank}$ profiles, we computed the radial profiles of the CLEANed images as well as the radial profiles of the CLEAN models. To maximize the angular resolution in the radial profiles, we used a $\mathtt{robust}$ parameter of 0.0, which yielded a beam size of $0.043^{\prime\prime} \times 0.043^{\prime\prime}$. To better enable a direct comparison between the 1.3 mm and 3 mm data (particularly for calculating the spectral index), we restricted the $uv$-range of CLEAN images to the intersection of the two datasets: 11.5 - 5,400 k$\lambda$. For our "CLEAN model" radial profiles, we compute the azimuthal average of the CLEAN components, without the residuals. The resulting profiles are shown in Figure \ref{fig:frankProfiles_CLEAN}. Notice the spectral index behavior in the gap and ring is consistent with that shown in Figure \ref{fig:frankProfiles} in both the CLEAN image azimuthal averages and the CLEAN Model azimuthal averages. Notably, there is structure inside the gap interior to and just outside the orbit of the proposed companion. 

We also recover all of the inner-disk substructure described in Section \ref{sec:substructure}, especially the gaps at 6 au and 14 au, which can be seen in the 1.3 mm CLEAN Model radial profile as well as the 3 mm profile. The wiggles in the outer disk of the 3mm $\mathtt{frank}$ radial profile are also seen in both the CLEAN and CLEAN Model radial profiles. The fact that they appear in the three radial profiles indicates that they are likely due to the very low (typically $\sim 1-2$) signal-to-noise ratio of the 3 mm data in the outer disk and are not simply Fourier artifacts in the $\mathtt{frank}$ fit.  

\begin{figure}[!htb]
    \centering
    \includegraphics[width=\linewidth]{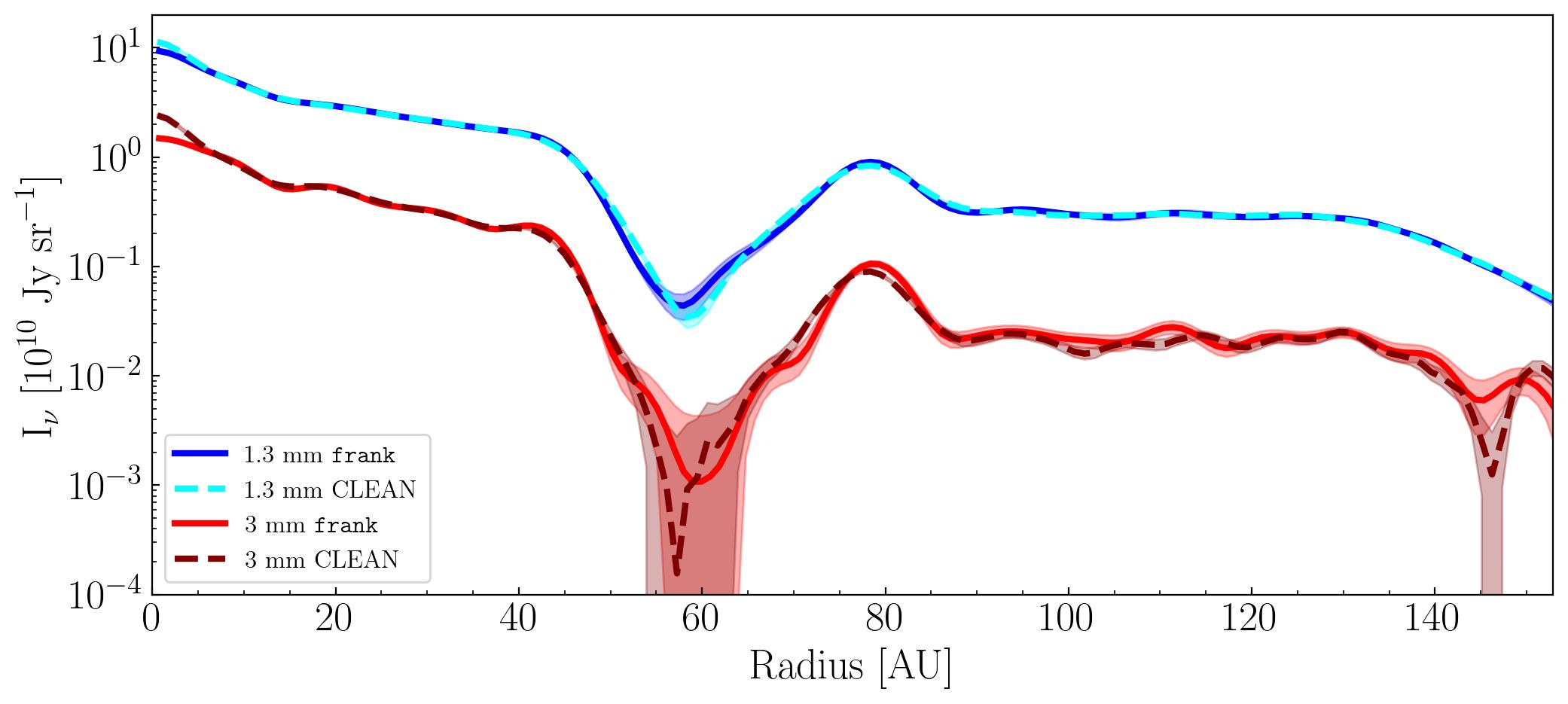}
    \caption{The 1.3 mm and 3 mm $\mathtt{frank}$ profiles (blue and red solid lines) convolved to the CLEAN beam resolution, compared with the CLEAN profiles (maroon and cyan dashed lines). The convolved $\mathtt{frank}$ profiles are consistent with the CLEAN profiles within the 1$\sigma$ uncertainties of the data. }
    \label{fig:frankCLEANBeam}
\end{figure}

We also show the $\mathtt{frank}$ profile convolved to the CLEAN beam in Figure \ref{fig:frankCLEANBeam} to ensure good agreement between the two. The CLEAN-beam-convolved $\mathtt{frank}$ profile matches the CLEAN image profile well. Critically for our SED analysis, the $\mathtt{frank}$ profiles are consistent with the flux levels of the CLEAN image profiles.

\begin{figure}[!htb]
    \centering
    \includegraphics[width=\linewidth]{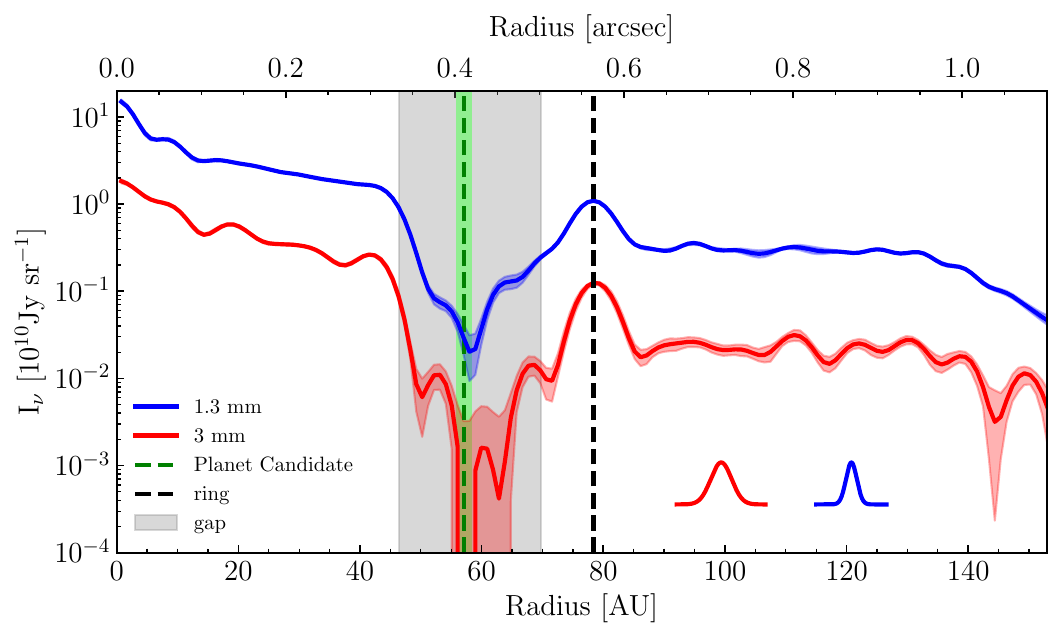}
    \caption{The 1.3 and 3 mm $\mathtt{frank}$ profiles shown at their effective resolutions of 15 and 29 mas respectively. Notice there are several features in the 1.3 mm $\mathtt{frank}$ profile that correspond well to features in the 3 mm profile but disappear in the convolution to 29 mas. The inset Gaussians show the effective resolution of the two profiles. }
    \label{fig:frankProfiles_nonConvolved}
\end{figure}

\section{Visibility Fits}
\begin{figure}[!htb]
    \centering
    \includegraphics[width=0.49\linewidth]{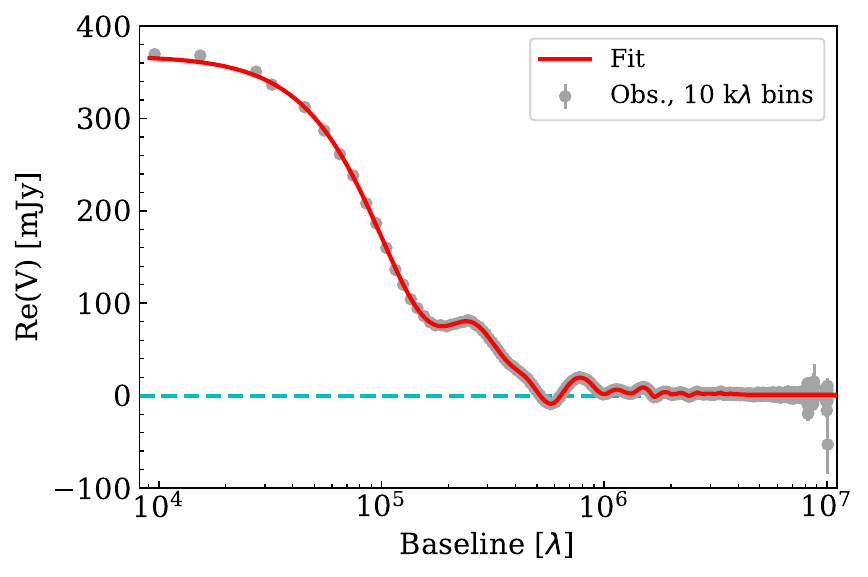}
    \includegraphics[width=0.49\linewidth]{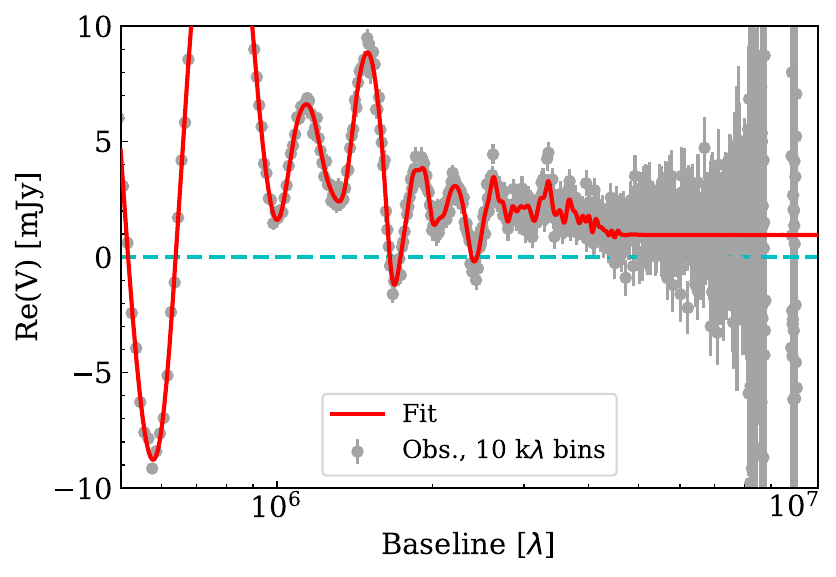}
    \caption{\textbf{Left:} The full range of visibilities in the 1.3 mm data, binned in 10 k$\lambda$ bins. \textbf{Right:} A zoom-in of the 1.3 mm visibilities, showing the longest baselines.}
    \label{fig:frank_1mm_vis}
\end{figure}

\begin{figure}[!htb]
    \centering
    \includegraphics[width=0.49\linewidth]{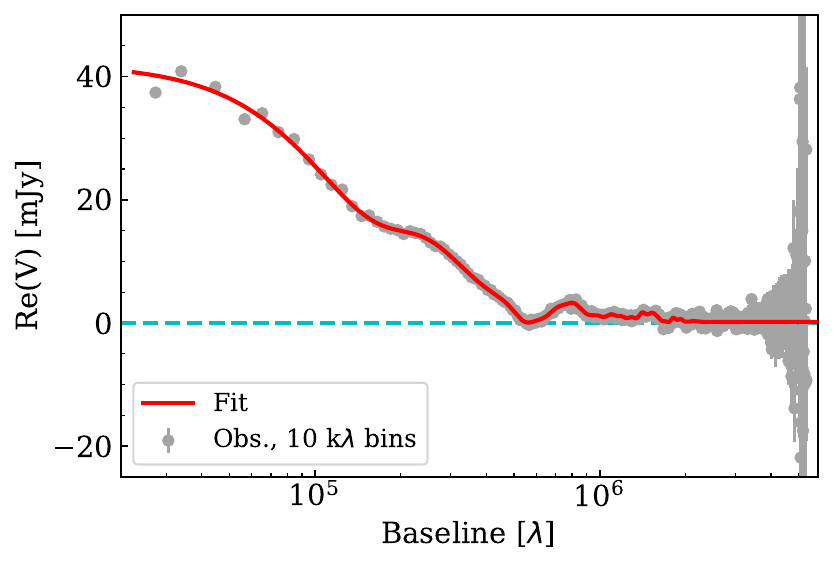}
    \includegraphics[width=0.49\linewidth]{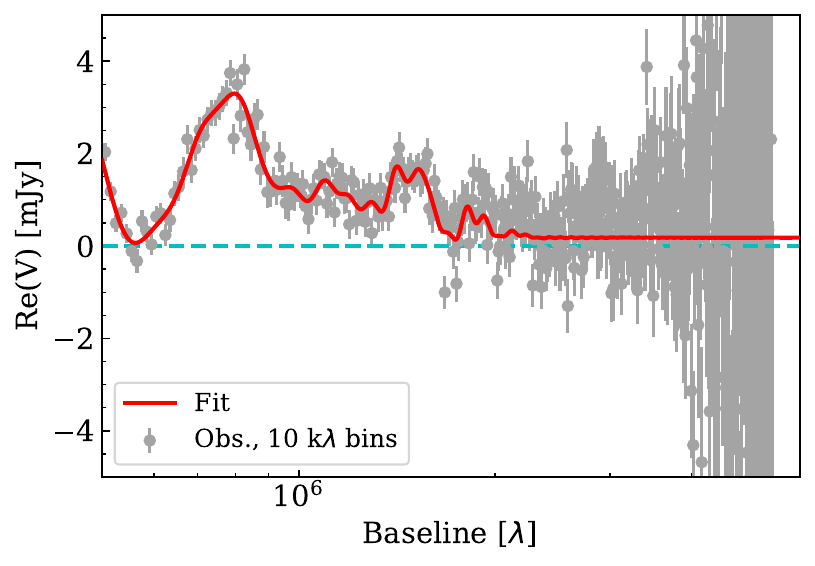}
    \caption{\textbf{Left:} The full range of visibilities in the 3 mm data, binned in 10 k$\lambda$ bins. \textbf{Right:} A zoom-in of the 3 mm visibilities, showing the longest baselines.}
    \label{fig:frank_3mm_vis}
\end{figure}

The $\mathtt{frank}$ procedure fits the visibility curves well, capturing all of the major features of the large and small scales. The curves and the corresponding $\mathtt{frank}$ fits are shown for the 1.3 mm band and 3 mm band in Figures \ref{fig:frank_1mm_vis} and \ref{fig:frank_3mm_vis}, respectively. 

\section{Point Source}
\begin{figure}[!htb]
    \centering
    \includegraphics[width=0.32\linewidth]{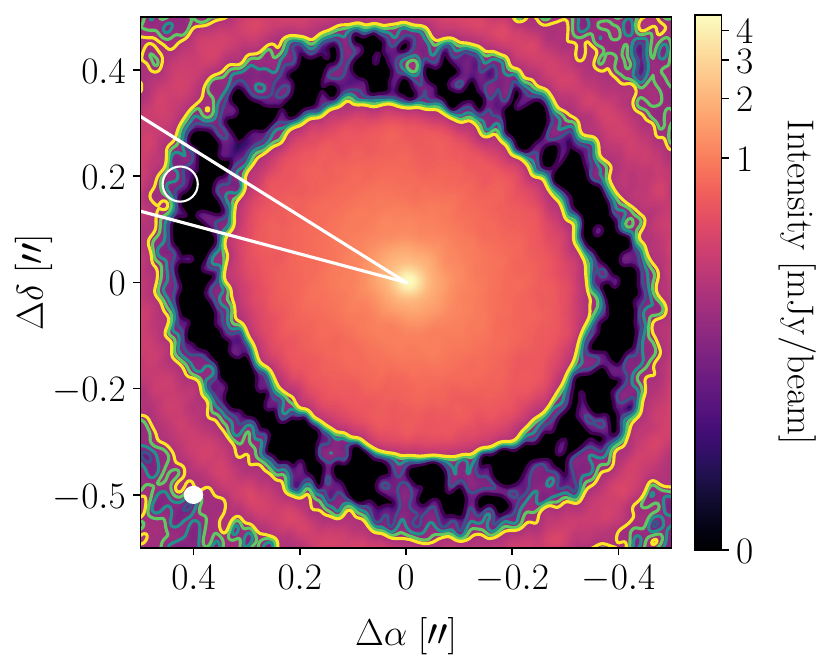}
    \includegraphics[width=0.32\linewidth]{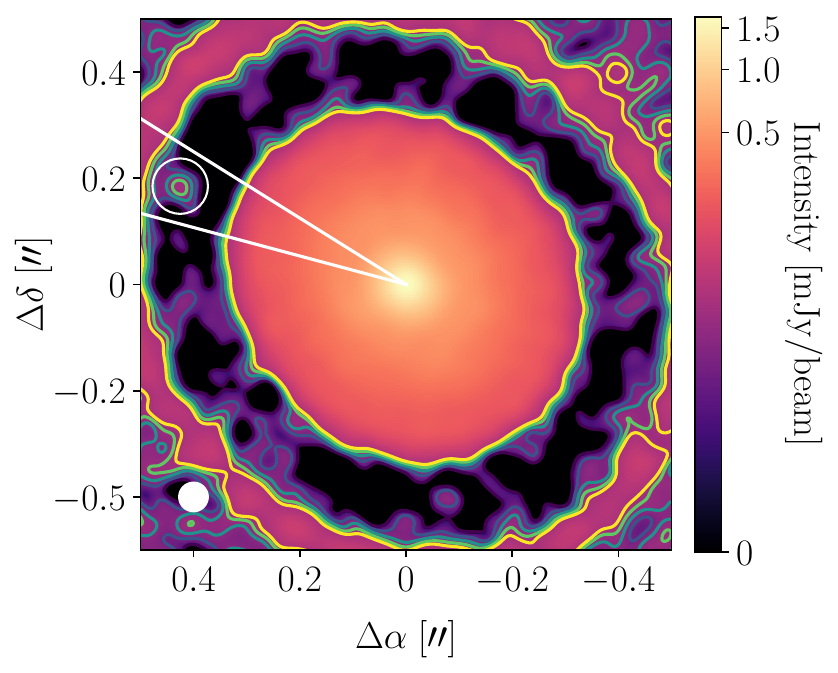}
    \includegraphics[width=0.32\linewidth]{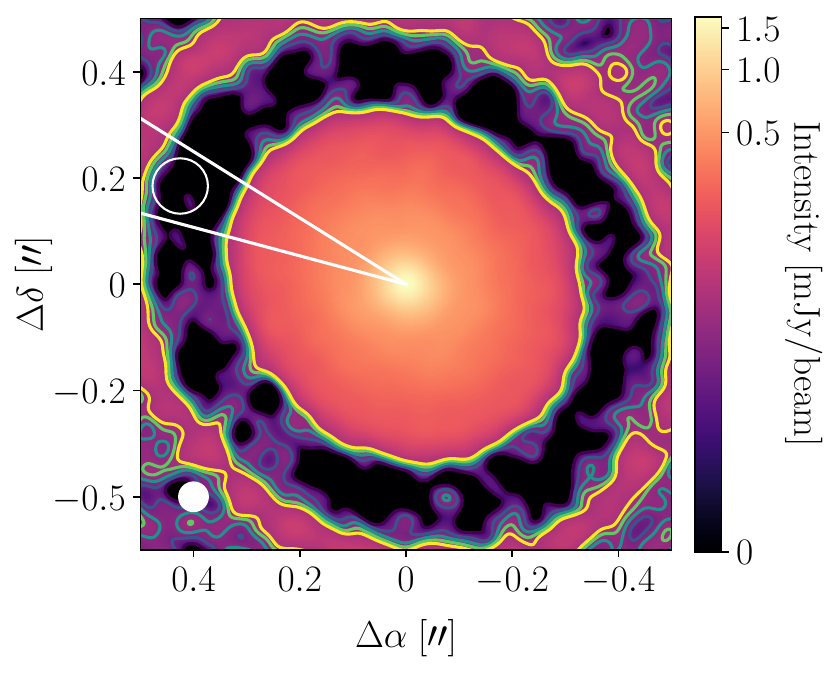}
    \caption{The CLEAN images, zoomed in to the disk center and gap, with a power law stretch applied to highlight to low-level emission. 1$\sigma$, 2$\sigma$, 3$\sigma$, 4$\sigma$, and 5$\sigma$ contours are plotted in each image. The white ellipse at the bottom left shows the CLEAN beam. The empty ring shows the location of the point source in the 3 mm image. \textbf{Left:} The 1.3 mm image, with the location of the point source highlighted. Notice there is no emission there, even at the 1$\sigma$ level. \textbf{Center:} The 3 mm image, with the point source location highlighted. The white lines mark the azimuth limits for the radial profile azimuthal averaging used in Appendix \ref{app:Tests}. \textbf{Right:} The 3 mm data with the point source subtracted from the visibilities, imaged using the same CLEAN parameters as the center panel. Notice the point source has been successfully removed from the image. }
    \label{fig:pointSource}
\end{figure}

We identify a point source in the 3 mm data that is not present in the 1.3 mm data. The point source has a peak flux of $\approx 5 \times 10^{-2}$ mJy/beam (a 5$\sigma$ detection in the 3 mm image) and a total integrated flux of $9 \times 10^{-2}$ mJy. To account for it, we follow a procedure similar to that used to remove the asymmetry in HD 143006 by \citet{Andrews_CPD_2021ApJ}. We identify the point source location in the CLEAN model, sample the image at the $u, v$ locations from our 3 mm Measurement Set, and subtract the resulting modeled visibilities from the 3 mm visibilities. The resulting before and after CLEAN images are shown in Figure \ref{fig:pointSource}

Using the 1.3 mm $RMS$ uncertainty ($2 \times 10^{-2}$ mJy/beam) as a lower bound for the flux in that band and the 3 mm peak flux, we estimate a spectral index of $\alpha = -1.1 \pm 0.3$ for the point source. The uncertainty in the spectral index is dominated by the 10\% overall flux uncertainty of the data. This is consistent with the spectral index of optically thick synchrotron emission from compact regions of Active Galactic Nuclei \citep[AGN;][]{Behar_AGN_2018MNRAS}.

The point source may then be a background millimeter galaxy, which is revealed by the proper motion of the disk moving it into the gap during the 3 mm observations, whereas during the 1.3 mm observations it may have been obscured by the disk. In the $\sim 20$ months between the 1.3 mm and 3 mm observations, the proper motion of Elias 24 was $\Delta \alpha \sim -13$ mas and $\Delta \delta \sim -41$ mas \citep{Gaia_catalogueDR3_2020yCat}, which would move the source from the edge of the inner disk, closer to the center of the gap. 

\section{Residuals} \label{app:residuals}

\begin{figure}[!htb]
    \centering
    \includegraphics[width=0.32\linewidth]{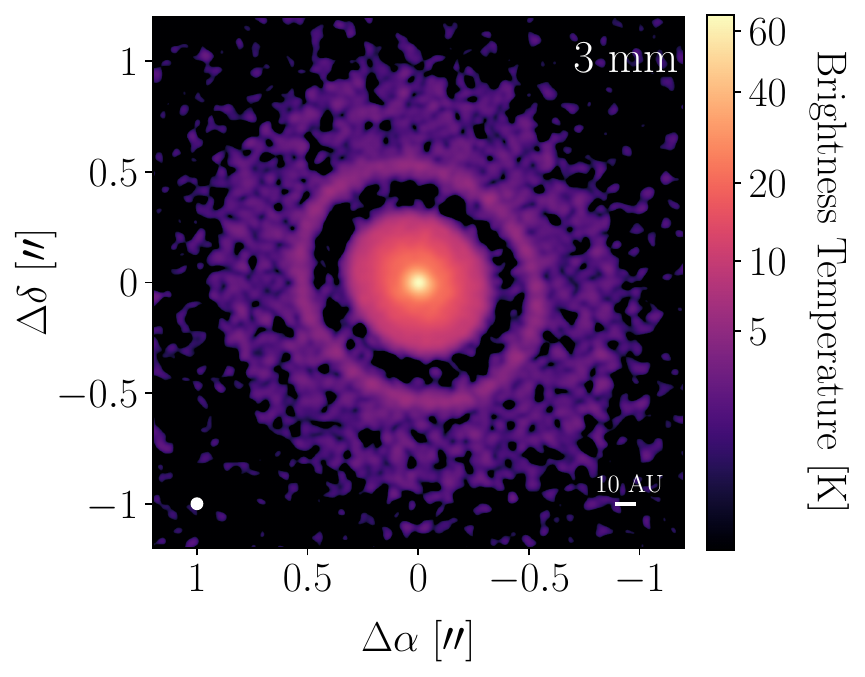}
    \includegraphics[width=0.32\linewidth]{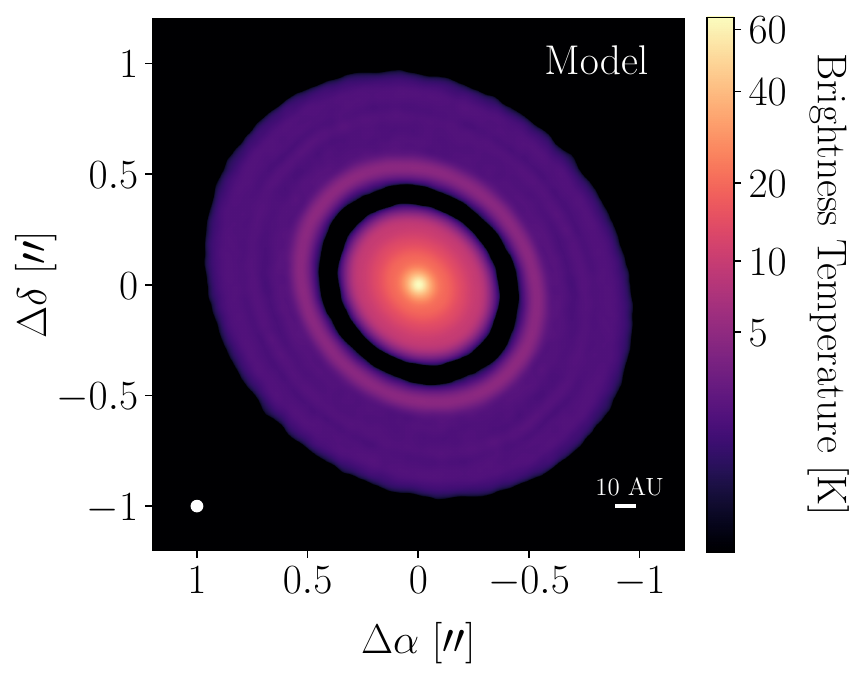}
    \includegraphics[width=0.32\linewidth]{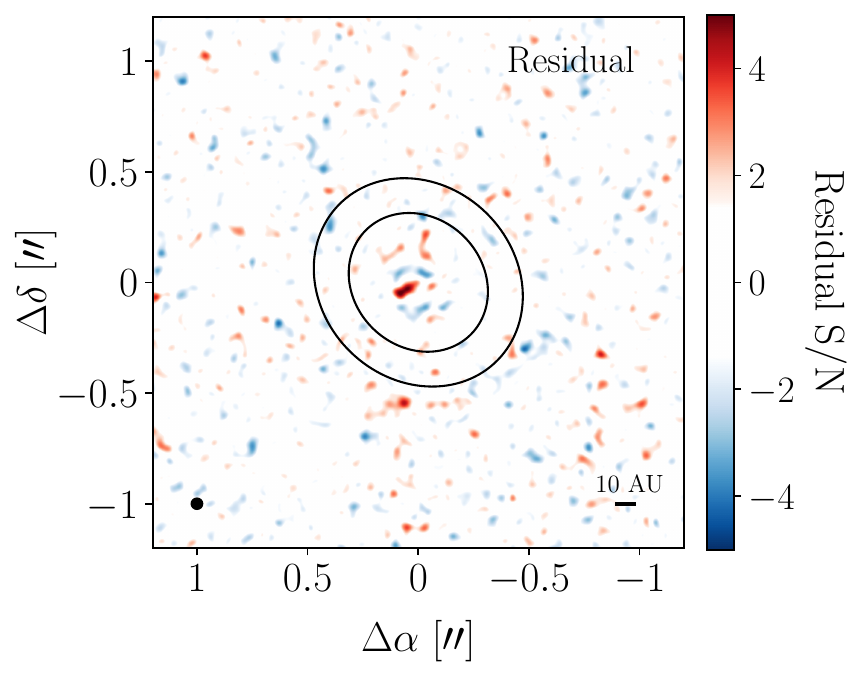}
    \caption{The CLEAN images of the point-source-subtracted 3 mm visibilities (left), the $\mathtt{frank}$ fit visibilities (middle), and the 3 mm visibilities - $\mathtt{frank}$ fit visibilities residuals (right). Notice all residuals remaining are at approximately the 1$\sigma$ level.}
    \label{fig:Residuals}
\end{figure}

We computed the residuals of the 3 mm $\mathtt{frank}$ fits by sampling the fit at the $(u, \ v)$ locations of the 3 mm data and subtracting that from the point-source-subtracted 3 mm visibilities. we then imaged the $\mathtt{frank}$ fit visibilities and the residuals using the same CLEAN parameters as the 3 mm data, and present the results in Figure \ref{fig:Residuals}. 

We do not see any contribution from the point source in the residuals, indicating our point source subtraction was successful and the $\mathtt{frank}$ fit is not contaminated by it. The residuals look like noise overall, with no significant structure beyond the 1$\sigma$ level.

\end{document}